\documentclass[aps,pra,onecolumn,amsmath,superscriptaddress,floatfix,notitlepage,showpacs]{revtex4-2}

\usepackage[OT1]{fontenc}

\usepackage{MnSymbol}
\usepackage{physics}

\usepackage{geometry}
\usepackage[table,xcdraw]{xcolor}
\usepackage{graphicx}
\usepackage{multirow}
\usepackage{array}
\usepackage{longtable}

\usepackage{enumitem}
\usepackage{dsfont}
\usepackage{bbold}
\usepackage[normalem]{ulem}
\usepackage{accents}

\usepackage[colorlinks=true, allcolors=blue]{hyperref}

\newcommand{\rhoexp}{\hat{\rho}_{\mathrm{exp}}}
\newcommand{\rhoideal}{\hat{\rho}_{\mathrm{theor}}}

\begin{document}

\title{Quantum tomography of ququart states of optical vector vortices based on measurements in mutually unbiased bases}

\author{D. D. Reshetnikov}
\affiliation{St. Petersburg State University, 7/9 Universitetskaya Nab., 199034 St. Petersburg, Russia}
\email{d.d.reshetnikov@gmail.com}
\author{E. V. Maliutina}
\affiliation{St. Petersburg State University, 7/9 Universitetskaya Nab., 199034 St. Petersburg, Russia}
\author{T. K. Korol}
\affiliation{St. Petersburg State University, 7/9 Universitetskaya Nab., 199034 St. Petersburg, Russia}
\author{E. A. Vashukevich}
\affiliation{St. Petersburg State University, 7/9 Universitetskaya Nab., 199034 St. Petersburg, Russia}
\author{T. Yu. Golubeva}
\affiliation{St. Petersburg State University, 7/9 Universitetskaya Nab., 199034 St. Petersburg, Russia}

\date{\today}

\begin{abstract}
In this paper we present an original method for quantum state tomography of vector optical vortices with an axially symmetric transverse polarization profile. We present an interference technique which allows one to distinguish states of vector vortices with a fixed topological charge with high accuracy. We also demonstrate the feasibility of experimental quantum tomography of the density matrix in the logical ququart space built on vector vortices, with high fidelity for all 20 basis states. It is shown that the proposed tomography scheme provides a uniform characterization of states in all mutually unbiased bases of the ququart space.

\end{abstract}

\maketitle

\section{Introduction}
\label{sec:intro}

Optical vector vortices (vector vortex beams, VVB) are structured light fields whose polarization and phase degrees of freedom form a hybrid four-dimensional state space~---~a ququart~\cite{AlKhafaji2022,Ndagano2018}. In the quantum optics formalism, the state of a VVB is described by a vector in the tensor product of the corresponding Hilbert spaces: a two-dimensional polarization state space and a multidimensional orbital angular momentum (OAM) space with different values of the topological charge $\ell_1, \ell_2$, $\mathcal{H} = \mathcal{H}_P \otimes \mathcal{H}_{\mathrm{OAM}=\ell_1, \ell_2}$. The subclass of VVB built on the OAM Hilbert space with a specific value of OAM $\mathcal{H}_{\mathrm{OAM}=\pm\ell}$ is of particular interest. Such states have an axially symmetric transverse polarization distribution. As shown in Ref.~\cite{Sokolov2017}, the polarization structure of these states is invariant under rotation of the transmitting and receiving equipment, which makes axially symmetric VVB promising for space communication systems~\cite{Reshetnikov2024}.

The hybrid polarization-phase nature of VVB provides additional possibilities for the use of such fields in quantum information applications. In quantum imaging, VVB are used to suppress scattering and improve image contrast. In particular, the partial preservation of the polarization structure during VVB propagation in a scattering medium enables their use in a quantum scanner for early cancer diagnosis~\cite{Gianani2020}. In optical manipulation, tightly focused VVB create forces and torques unreachable for scalar beams, which allows multiplexed trapping and rotation of microparticles, as shown in Ref.~\cite{Zhang2023,Li2026}. Optical vector vortices also find application in quantum metrology, where the axial symmetry of the fields improves the accuracy of angular-displacement measurements, as demonstrated in Ref.~\cite{Yuan2019,Balzarotti2017}. According to Ref.~\cite{Ndagano2018}, in high-dimensional optical communication, hybrid polarization-OAM encoding increases the information capacity of the channel and improves the resilience of information transmission to noise. The most mature quantum application of VVB is the use of such beams in quantum key distribution (QKD) systems~\cite{Rosales-Guzmán2018}.

The application of VVB in QKD protocols has proceeded in two directions. The first involves the construction of qudit protocols based on entangled Bell states in mutually unbiased bases (MUB). The deterministic detector scheme for vector vortex states proposed in Ref.~\cite{Deterministic2017} ensures a detection probability of unity for Bell states with a low QBER. Following Refs.~\cite{Reshetnikov2024, Reshetnikov2026}, the second direction relies on the rotational invariance of VVB to eliminate the need for polarization-basis alignment between Alice and Bob. The high dimension of the Hilbert space of vector vortex states makes it possible to increase the secret key generation rate in QKD protocols, as well as to improve their noise resilience and cryptographic security compared with qubit-based protocols~\cite{Mirhosseini2015, Sit2017, Kulik2007}.

The purity $\mathcal{P}$ and fidelity $\mathcal{F}$ of the generated states directly determine the performance of QKD protocols. Purity characterizes the degree of mixedness of a state, which can be caused both by imperfect state preparation and by correlations with the environment that a potential eavesdropper can exploit~\cite{InformationComplementarity2015,MixedStatesQKD2020}. Fidelity shows how close the reconstructed state is to the target state and determines the probability of correct detection~\cite{Czerwinski2022}. As shown in Ref.~\cite{Parigi2015}, for VVB, purity and fidelity remain at a high level under changes in the receiver orientation. Thus, control over the purity and fidelity of the generated states is a necessary condition for the security and performance of QKD protocols.

The purity and fidelity of quantum states can be calculated using quantum tomography~\cite{DAriano2003,Paris2004}. This procedure reconstructs the density matrix from the results of measurements on an ensemble of identically prepared systems. For a quantum system of dimension $d$, standard tomography requires the measurement of $d^2-1$ physical quantities, which leads to $2d^2-d$ measurement outcomes~\cite{Czerwinski2022}. This number grows rapidly as the dimension increases, which motivates the development of more efficient approaches to density matrix reconstruction. MUB-based tomography reduces the required number of outcomes to $d^2+d$ and  provides an advantage over standard tomography for systems of dimension $d\geq3$, while ensuring minimal statistical error~\cite{Durt2010}. The existence of a complete set of MUB is proven for dimensions that are integer powers of a prime number. According to Ref.~\cite{DiazGuevara2021}, for composite dimensions, however, the question remains open. As shown in~\cite{DiazGuevara2021,Nagali2010}, for ququarts, the systems of interest in this article, 5 MUB are sufficient for constructing an informationally complete set of measurements.

Existing methods for VVB state tomography fall into three groups. The first group is based on Stokes projection measurements~\cite{Selyem2019}. This method allows the reconstruction of the density matrix through sequential analysis of the polarization and the spatial mode, but it requires considerable measurement time and is sensitive to drift in the setup parameters. The second group of methods uses generalized POVM measurements~\cite{AlKhafaji2022}. This method enables characterization of the vector mode in a single measurement, but it does not allow reconstruction the full density matrix of an arbitrary ququart state without prior knowledge of the basis. The third group, proposed in Ref.~\cite{Nagali2010}, is based on the use of q-plates for preparing and measuring hybrid ququarts in all 5 MUB. This method makes it possible to perform full tomography, but it suffers from a significant drawback: the q-plates fix the topological charge and the operating wavelength during fabrication, which limits the universality of the setup~\cite{QPlateLimitations2023}. An alternative method for vector vortex mode tomography based on polarimetry and analysis of transverse intensity profiles has recently been proposed in Ref.~\cite{Braga2026}. However, its information completeness has been established only for first-order modes ($\ell=\pm 1$), and the detection method based on a CCD camera requires long signal accumulation in the few-photon regime.
 
In this paper we propose a method for quantum tomography of ququart states based on optical vector vortices within the MUB framework. The measurements of the polarization-phase structure of the states are based on a cascade of polarization Mach-Zehnder interferometers and phase holograms displayed on spatial light modulators. We propose a procedure for reconstructing the density matrix of an unknown ququart state from a set of 15 projective measurements in 5 MUB. The paper is structured as follows. Section~\ref{sec:methods} describes the experimental implementation of the procedure for preparing and measuring ququart states of optical vector vortices, as well as the procedure for density matrix reconstruction. Section~\ref{sec:results} presents the results of full tomography of all 20 basis states, including fidelity and purity values for the raw and processed data. Section~\ref{sec:discussion} discusses the role of the physical projective procedure and the contributions of individual error sources. It also compares the obtained results with the published data. The main conclusions are summarized in Section~\ref{sec:conclusion}. Additional materials, including the full density matrices for all 20 states and their graphical comparison with the theoretical values, are given in the Supplementary Materials.

\section{Methods}
\label{sec:methods}
\subsection{Construction of a complete MUB set}

Full tomography of a ququart state requires projective measurements in a complete set of mutually unbiased bases. Two orthonormal bases \(\{|e_i\rangle\}\) and \(\{|f_j\rangle\}\) are called mutually unbiased if the condition \(|\langle e_i|f_j\rangle|^2 = 1/d\) holds for any pair of vectors from different bases, where \(d\) is the dimension of the Hilbert space~\cite{Durt2010}. This condition means that a measurement in one basis yields no information about the outcome of a measurement in another basis, which ensures minimal statistical error in density matrix reconstruction. The maximum number of mutually unbiased bases in a space of dimension \(d\) equals \(d+1\). Thus, for ququarts, the complete set consists of 5 MUB each containing 4 orthonormal states.

The general construction of a complete set of MUB is based on the eigenstates of the operators of the Heisenberg-Weyl group~\cite{Durt2010,Bengtsson2017}. On a \(d\)-dimensional logical state space, one can define the generalized shift operator \(X\) and the phase operator \(Z\):
\begin{equation}
X|j\rangle = |j+1 \bmod d\rangle, \qquad Z|j\rangle = \omega^j|j\rangle,
\label{eq:weyl}
\end{equation}
where \(\omega = e^{2\pi i/d}\), and \(|j\rangle\), \(j = 0,\ldots,d-1\), is the logical basis of this space. The operators \(X\) and \(Z\) satisfy the relation \(ZX = \omega XZ\). Thus, following~\cite{Durt2010,Bengtsson2017}, for \(d = 4\), a complete set of 5 MUB can be constructed as the sets of eigenvectors of the operators \(Z\), \(X\), \(XZ\), \(XZ^2\), and \(XZ^3\).

\begin{figure}[htb]
     \centering
     \includegraphics[width=0.85\columnwidth]{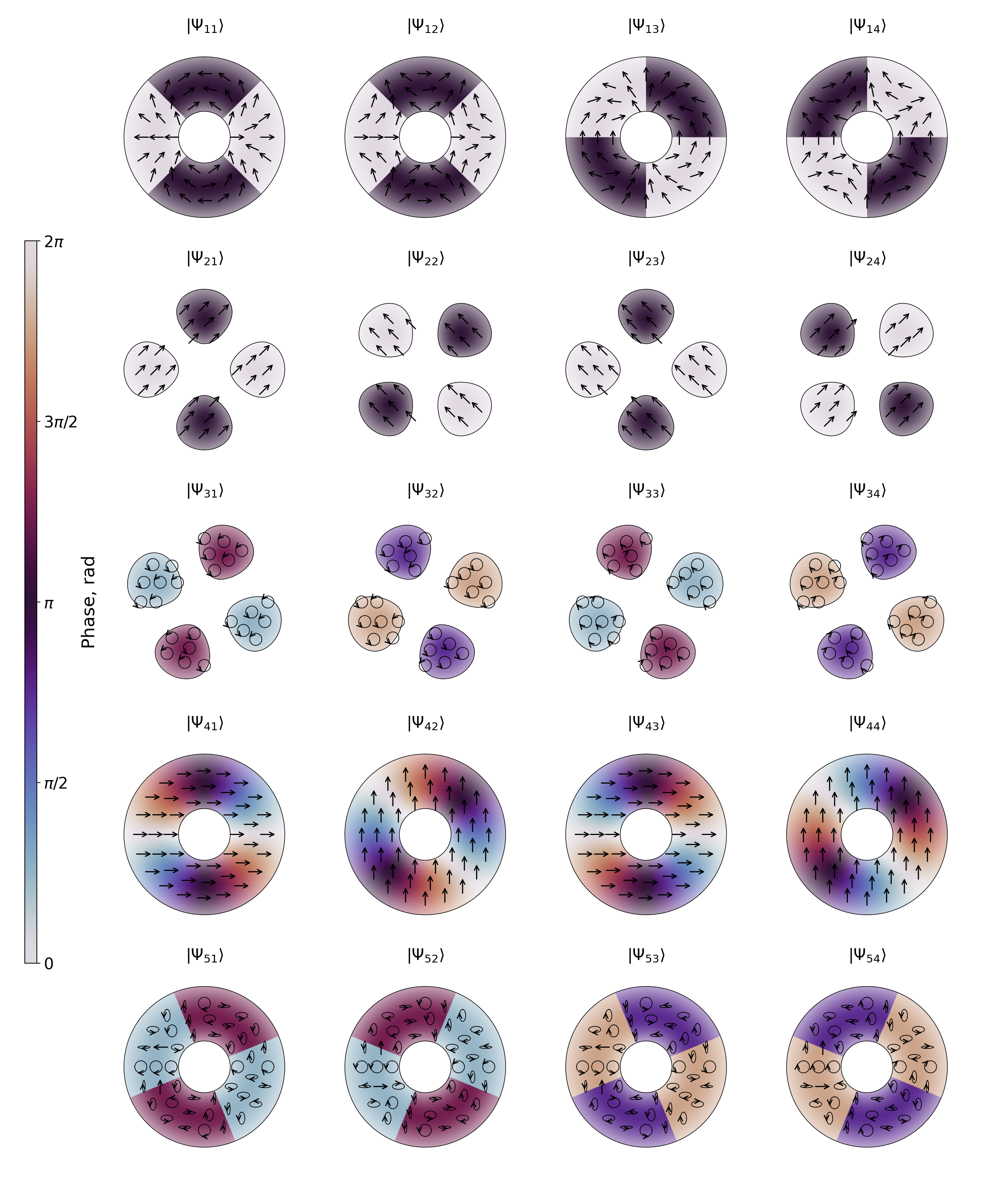}
     \caption{Intensity profiles (the contour is marked by a solid line), phases (shown in color), and polarization vectors (linear polarization states are marked by arrows; circles and ellipses denote circular and elliptical polarization states, respectively) for all 20 MUB states.}
     \label{Fig: MUBtates}
     \end{figure}

We consider a vector vortex field representable as the tensor product of a superposition of two orthogonal polarization modes and a superposition of two spatial modes with a fixed OAM magnitude. The state space of such a field can naturally be described as a ququart logical space:
\begin{equation}
	|0\rangle = |H, +\ell\rangle,\quad |1\rangle = |H, -\ell\rangle,\quad |2\rangle = |V, +\ell\rangle,\quad |3\rangle = |V, -\ell\rangle.
	\label{eq:logicalPI}
\end{equation}
where the index $\pm\ell$ denotes the order of the phase singularity, that is, the number of phase jumps during a full turn in the transverse plane around the propagation axis. The ket vectors $\ket{H}, \ket{V}$ denote states with horizontal and vertical polarizations, and $\ket{\pm\ell}$ are spatial modes with OAM $\pm\ell$. It should be noted that, when considering the entire set of modes with different values of the OAM, the state space can also be decomposed into a set of ququart subspaces with a fixed absolute value of the OAM. In this paper, we consider VVB states with OAM $\pm 2$, because they are convenient to generate. The 20 basis states belonging to the 5 MUB for ququart tomography are constructed following~\cite{Nagali2010} and are listed in Table~\ref{tab: mub-states}. The intensity and phase profiles and the polarization-vector distribution in the transverse plane for the 20 basis states belonging to the 5 MUB are shown in Fig.~\ref{Fig: MUBtates}.
\begin{table}[h!]
	\caption{The set of states used for MUB tomography of a ququart. The partitions correspond to the 5 MUB (I--V), and the rows correspond to the 4 states within each basis. Each state is given as an expansion in the basis of scalar optical vortices according to Eq.~\eqref{eq:logicalPI}.}
	\begin{tabular}{|c|c|l|}
		\hline
		MUB set            & State           & Vector in the computational basis                                 \\ \hline
		\multirow{4}{*}{I}   & $|\Psi_{11}\rangle$ & $\frac{1}{2}|0\rangle + \frac{1}{2}|1\rangle - \frac{i}{2}|2\rangle + \frac{i}{2}|3\rangle$  \\ \cline{2-3} 
		& $|\Psi_{12}\rangle$ & $\frac{1}{2}|0\rangle + \frac{1}{2}|1\rangle + \frac{i}{2}|2\rangle - \frac{i}{2}|3\rangle$  \\ \cline{2-3} 
		& $|\Psi_{13}\rangle$ & $\frac{i}{2}|0\rangle - \frac{i}{2}|1\rangle + \frac{1}{2}|2\rangle + \frac{1}{2}|3\rangle$  \\ \cline{2-3} 
		& $|\Psi_{14}\rangle$ & $-\frac{i}{2}|0\rangle + \frac{i}{2}|1\rangle + \frac{1}{2}|2\rangle + \frac{1}{2}|3\rangle$ \\ \hline
		\multirow{4}{*}{II}  & $|\Psi_{21}\rangle$ & $\frac{1}{2}|0\rangle + \frac{1}{2}|1\rangle + \frac{1}{2}|2\rangle + \frac{1}{2}|3\rangle$  \\ \cline{2-3} 
		& $|\Psi_{22}\rangle$ & $\frac{i}{2}|0\rangle - \frac{i}{2}|1\rangle - \frac{i}{2}|2\rangle + \frac{i}{2}|3\rangle$  \\ \cline{2-3} 
		& $|\Psi_{23}\rangle$ & $\frac{1}{2}|0\rangle + \frac{1}{2}|1\rangle - \frac{1}{2}|2\rangle - \frac{1}{2}|3\rangle$  \\ \cline{2-3} 
		& $|\Psi_{24}\rangle$ & $-\frac{i}{2}|0\rangle + \frac{i}{2}|1\rangle - \frac{i}{2}|2\rangle + \frac{i}{2}|3\rangle$ \\ \hline
		\multirow{4}{*}{III} & $|\Psi_{31}\rangle$ & $\frac{i}{2}|0\rangle + \frac{1}{2}|1\rangle - \frac{1}{2}|2\rangle + \frac{i}{2}|3\rangle$  \\ \cline{2-3} 
		& $|\Psi_{32}\rangle$ & $-\frac{i}{2}|0\rangle + \frac{1}{2}|1\rangle + \frac{1}{2}|2\rangle + \frac{i}{2}|3\rangle$ \\ \cline{2-3} 
		& $|\Psi_{33}\rangle$ & $\frac{1}{2}|0\rangle + \frac{i}{2}|1\rangle - \frac{i}{2}|2\rangle + \frac{1}{2}|3\rangle$  \\ \cline{2-3} 
		& $|\Psi_{34}\rangle$ & $\frac{1}{2}|0\rangle - \frac{i}{2}|1\rangle - \frac{i}{2}|2\rangle - \frac{1}{2}|3\rangle$  \\ \hline
		\multirow{4}{*}{IV}  & $|\Psi_{41}\rangle$ & $1|0\rangle + 0|1\rangle + 0|2\rangle + 0|3\rangle$                                          \\ \cline{2-3} 
		& $|\Psi_{42}\rangle$ & $0|0\rangle + 0|1\rangle + i|2\rangle + 0|3\rangle$                                          \\ \cline{2-3} 
		& $|\Psi_{43}\rangle$ & $0|0\rangle + 1|1\rangle + 0|2\rangle + 0|3\rangle$                                          \\ \cline{2-3} 
		& $|\Psi_{44}\rangle$ & $0|0\rangle + 0|1\rangle + 0|2\rangle - i|3\rangle$                                          \\ \hline
		\multirow{4}{*}{V}   & $|\Psi_{51}\rangle$ & $\frac{i}{2}|0\rangle + \frac{1}{2}|1\rangle - \frac{i}{2}|2\rangle + \frac{1}{2}|3\rangle$  \\ \cline{2-3} 
		& $|\Psi_{52}\rangle$ & $\frac{1}{2}|0\rangle + \frac{i}{2}|1\rangle - \frac{1}{2}|2\rangle + \frac{i}{2}|3\rangle$  \\ \cline{2-3} 
		& $|\Psi_{53}\rangle$ & $\frac{1}{2}|0\rangle - \frac{i}{2}|1\rangle + \frac{1}{2}|2\rangle + \frac{i}{2}|3\rangle$  \\ \cline{2-3} 
		& $|\Psi_{54}\rangle$ & $-\frac{i}{2}|0\rangle + \frac{1}{2}|1\rangle - \frac{i}{2}|2\rangle - \frac{1}{2}|3\rangle$ \\ \hline
	\end{tabular}
	\label{tab: mub-states}
\end{table}

\subsection{State preparation}

\begin{figure}[htbp]
     \centering
     \includegraphics[width=0.6\columnwidth]{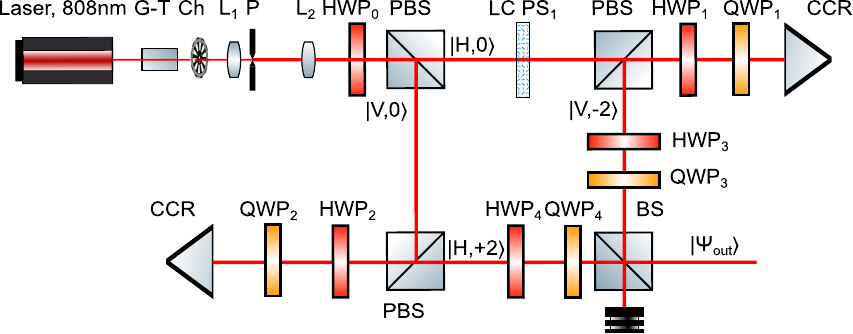}
     \caption{Optical scheme for preparing optical vector vortex states in bases I, II, III, and V. $Laser$~---~solid-state laser, $\lambda=808$~nm, $G-T$~---~Glan-Thompson prism, $Ch$~---~optical chopper wheel, $L_1, L_2$~---~converging lenses, $P$~---~pinhole, $HWP$~---~half-wave plate, $QWP$~---~quarter-wave plate, $BS$~---~beam-splitting cube, $PBS$~---~polarizing beam-splitting cube, $LC\ PS_1$~---~liquid-crystal variable retarder, $CCR$~---~cubic corner reflector. The ket vectors denote the polarization and OAM states in the scheme.}
     \label{Fig: preparation1}
     \end{figure}

We now consider the experimental implementation of a generator of few-photon states of quantum axially symmetric VVB. The experimental setup (Fig.~\ref{Fig: preparation1}) is based on a semiconductor laser with a wavelength of $\lambda=808$~nm (Nolatech FPL-808).
The light is coupled out of a single-mode fiber and collimated by a converging lens (not shown in the figure). A Glan-Thompson prism (ThorLabs GTH5-B) is used to ensure a high degree of polarization purity of the input beam and produce linear polarization parallel to the plane of the optical table. Subsequent spatial filtering is performed by an astigmatic telescopic system of two converging lenses with focal lengths $F_1=50$~mm and $F_2=300$~mm, mounted at a distance $F_1+F_2$ from each other. A pinhole aperture at the beam waist suppresses higher-order spatial modes and selects the fundamental Gaussian mode. The pinhole diameter is selected experimentally to maximize suppression of higher modes while preserving the intensity of the fundamental mode. The output is a Gaussian beam with horizontal polarization and a diameter of $8.0\pm0.2$~mm. The few-photon generation regime is achieved using a rotating optical chopper operating at 1~kHz and a symmetric pulse shape. The required attenuation level, corresponding on average to 0.1 photon per correlator bin, is achieved through a combination of neutral-density (ND) filters and selection of the exposure time of the coincidence circuit.

We use a polarization Mach-Zehnder interferometer with cubic corner reflectors (CCR) with an interference phase-shifting coating in both arms. A half-wave plate $HWP_0$ (ThorLabs WPH10M-808) redistributes the intensity of the input radiation between the interferometer arms. The half-wave plates $HWP_{1,2}$ and the quarter-wave plates $QWP_{1,2}$ (ThorLabs WPQ10M-808) are oriented at fixed angles of $22.5^\circ$ and $45^\circ$, respectively. They convert linear polarization into circular polarization before reflection from the CCR. A key feature of the CCRs used in our setup is that three total internal reflections at mutually orthogonal faces reverse the direction of circular polarization and produces a second-order scalar optical vortex (a vortex with a spatially uniform polarization state). The action of such an element on a Gaussian mode with circular polarization is described as follows:
	\begin{equation}
		\ket{R, 0} \to \ket{L, -2}, \qquad \ket{L, 0} \to \ket{R, +2}.
		\label{eq:ccr}
	\end{equation}
A detailed description of the polarization transformations in the CCR interferometer, based on the Jones matrix formalism, is presented in Ref.~\cite{Reshetnikov2025, Reshetnikov2026}.

According to Eq.~\eqref{eq:ccr}, the state $\ket{V,-2}$ is formed in the upper arm after the polarizing beam splitter (PBS), and the state $\ket{H,+2}$ is formed in the lower arm. The required vector vortex state results from constructive interference of these components at the output beam splitter of the interferometer, and the relative phase between the superposition terms is set by the angles of $HWP_{3,4}$ and $QWP_{3,4}$ in each arm before the output BS. Due to the speciality of spatial profile of the states produced by the CCR~\cite{Reshetnikov2025}, a vector vortex state is also formed in the second output channel of this BS as a result of destructive interference. However, these states are not considered further.

The plate orientation angles for each of the 16 states of bases I--III and V are determined numerically by optimizing the Jones matrices that describe the action of the sequence of polarization elements on the photon state. The results are listed in Table~II. In addition, the mutual orientation angle of the CCRs is varied depending on the required spatial mode profile of the vector vortex. The mutual CCR orientation requires retuning when passing from one state to another because the radiation reflected from an individual corner reflector has a non-axisymmetric distribution. A six-sector structure appears in the near diffraction zone, and a six-spot structure appears in the far zone. To form a correct intensity distribution at the interferometer output, one needs, in addition to phase tuning, to orient the CCRs so that the sector structures in both arms coincide in phase and position.

\begin{figure}[htbp]
     \centering
     \includegraphics[width=\columnwidth]{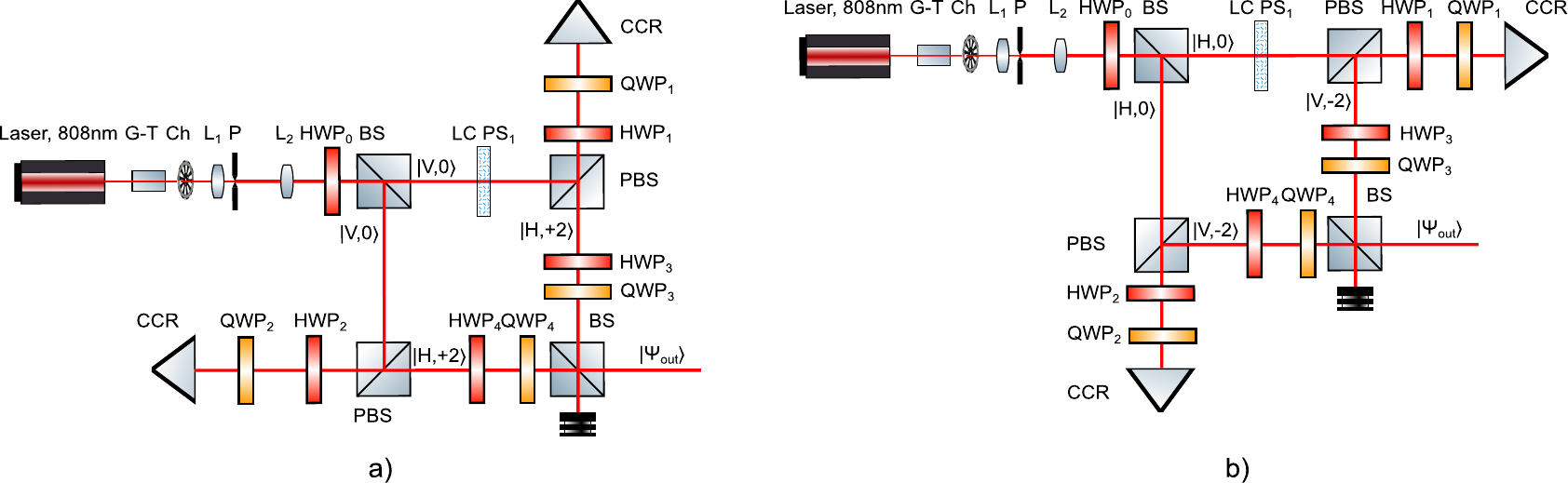}
     \caption{Optical scheme for preparing optical vector vortex states a) $\ket{\psi_{41}}, \ket{\psi_{42}}$, b) $\ket{\psi_{43}}, \ket{\psi_{44}}$. $Laser$~---~solid-state laser, $\lambda=808$~nm, $G-T$~---~Glan-Thompson prism, $Ch$~---~optical chopper wheel, $L_1, L_2$~---~converging lenses, $P$~---~pinhole, $HWP$~---~half-wave plate, $QWP$~---~quarter-wave plate, $BS$~---~beam-splitting cube, $PBS$~---~polarizing beam-splitting cube, $LC\ PS_1$~---~liquid-crystal variable retarder, $CCR$~---~cubic corner reflector. The ket vectors denote the polarization and OAM states in the scheme.}
     \label{Fig: preparation4}
     \end{figure}

It is worth noting that the proposed scheme is not universal and does not allow states with the same OAM to be formed in both arms of the interferometer simultaneously. Therefore, forming scalar optical vortex states from basis IV (the computational basis) required a minor modification of the scheme. The first beam-splitting cube of the interferometer was changed from a polarizing one to a conventional one, $PBS\rightarrow BS$ (see Fig.~\ref{Fig: preparation4}a,~\ref{Fig: preparation4}b).
In this case, a state with the same polarization and OAM is formed in both arms of the interferometer. The CCRs are rotated relative to each other by an angle of $30^{\circ}$ so that a structure of 12 overlapping sectors is formed in the near diffraction zone. In the far-field diffraction zone, this structure transforms into a scalar optical vortex state with a continuous phase distribution. The scalar polarization state in each arm is set by the orientation of the half-wave plates $HWP_{3,4}$ according to Table~\ref{tab: wave-plates}.

\begin{table}[h!]
	\caption{Orientation angles of the wave plates that provide preparation of each of the 20 MUB states.}
	\begin{tabular}{|c|c|c|c|c|c|c|}
		\hline
		MUB set            & State           & $\mathrm{HWP}_0$, $^\circ$ & $\mathrm{HWP}_3$, $^\circ$ & $\mathrm{QWP}_3$, $^\circ$ & $\mathrm{HWP}_4$, $^\circ$ & $\mathrm{QWP}_4$, $^\circ$ \\ \hline
		\multirow{4}{*}{I}   & $|\Psi_{11}\rangle$ & \multirow{12}{*}{22.5}     & -22.5                      & 0                          & 67.5                       & 0                          \\ \cline{2-2} \cline{4-7} 
		& $|\Psi_{12}\rangle$ &                            & 22.5                       & 0                          & 22.5                       & 0                          \\ \cline{2-2} \cline{4-7} 
		& $|\Psi_{13}\rangle$ &                            & 22.5                       & 90                         & -67.5                      & -90                        \\ \cline{2-2} \cline{4-7} 
		& $|\Psi_{14}\rangle$ &                            & 67.5                       & 90                         & 67.5                       & 90                         \\ \cline{1-2} \cline{4-7} 
		\multirow{4}{*}{II}  & $|\Psi_{21}\rangle$ &                            & 22.5                       & 45                         & 67.5                       & 90                         \\ \cline{2-2} \cline{4-7} 
		& $|\Psi_{22}\rangle$ &                            & 22.5                       & 45                         & 112.5                      & 45                         \\ \cline{2-2} \cline{4-7} 
		& $|\Psi_{23}\rangle$ &                            & -22.5                      & -45                        & 22.5                       & -45                        \\ \cline{2-2} \cline{4-7} 
		& $|\Psi_{24}\rangle$ &                            & -67.5                      & -45                        & -112.5                     & -45                        \\ \cline{1-2} \cline{4-7} 
		\multirow{4}{*}{III} & $|\Psi_{31}\rangle$ &                            & -22.5                      & -90                        & 67.5                       & 0                          \\ \cline{2-2} \cline{4-7} 
		& $|\Psi_{32}\rangle$ &                            & 67.5                       & 90                         & 67.5                       & 0                          \\ \cline{2-2} \cline{4-7} 
		& $|\Psi_{33}\rangle$ &                            & -22.5                      & 0                          & 67.5                       & 90                         \\ \cline{2-2} \cline{4-7} 
		& $|\Psi_{34}\rangle$ &                            & -22.5                      & 0                          & -22.5                      & -90                        \\ \hline
		\multirow{4}{*}{IV}  & $|\Psi_{41}\rangle$ & 45                         & 0                          & 0                          & 0                          & 0                          \\ \cline{2-7} 
		& $|\Psi_{42}\rangle$ & 45                         & 45                         & 0                          & 0                          & 0                          \\ \cline{2-7} 
		& $|\Psi_{43}\rangle$ & 0                          & 0                          & 0                          & 45                         & 0                          \\ \cline{2-7} 
		& $|\Psi_{44}\rangle$ & 0                          & 0                          & 0                          & 0                          & 0                          \\ \hline
		\multirow{4}{*}{V}   & $|\Psi_{51}\rangle$ & \multirow{4}{*}{22.5}      & -22.5                      & 45                         & 67.5                       & 45                         \\ \cline{2-2} \cline{4-7} 
		& $|\Psi_{52}\rangle$ &                            & -22.5                      & -45                        & -112.5                     & -45                        \\ \cline{2-2} \cline{4-7} 
		& $|\Psi_{53}\rangle$ &                            & 22.5                       & 45                         & 112.5                      & 45                         \\ \cline{2-2} \cline{4-7} 
		& $|\Psi_{54}\rangle$ &                            & -67.5                      & -45                        & 22.5                       & -45                        \\ \hline
	\end{tabular}
	\label{tab: wave-plates}
\end{table}
The schemes (Fig.~\ref{Fig: preparation1},~\ref{Fig: preparation4}) are tuned to a broadband interference regime. After the output beam splitter, the optical beams propagate coaxially and collinearly, and the intensity distribution exhibits a maximum at one output and a minimum at the other. This configuration provides high stability against longitudinal fluctuations of the optical arm lengths. To compensate for the slow drift of the relative phase caused by temperature fluctuations, we use a liquid-crystal variable retarder $LC\ PS_1$ LCC1513-B (ThorLabs) controlled by an applied voltage. In the absence of vibrations of the optical table, the relative phase drift does not exceed $\pi/20$ over tens of minutes, and this drift has no noticeable effect on the interference result. Fluctuations of the polarization in the initial laser mode have a substantially stronger effect on the quality of the formed states, which makes the Glan-Thompson prism a critically important element of the scheme.

\subsection{State measurement}

\label{sec:measurement}
\begin{figure}[htbp]
     \centering
     \includegraphics[width=0.65\columnwidth]{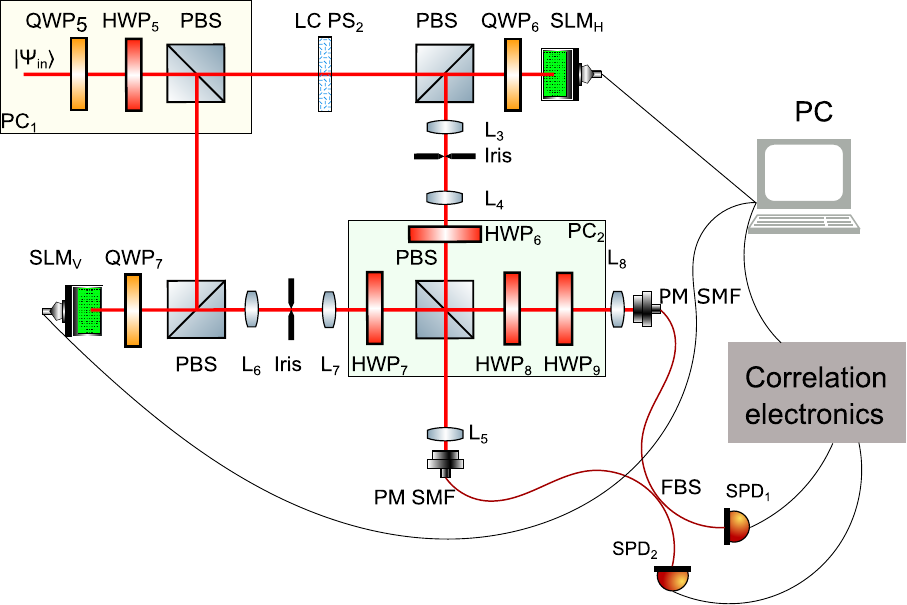}
     \caption{Optical scheme for ququart state tomography. $HWP$~---~half-wave plate, $QWP$~---~quarter-wave plate, $PBS$~---~polarizing beam-splitting cube, $LC\ PS_2$~---~liquid-crystal variable retarder, $SLM$~---~spatial light modulator, $L$~---~converging lens, $Iris$~---~iris diaphragm, $PM\ SMF$~---~polarization-maintaining single-mode fiber, $FBS$~---~fiber symmetric beam splitter, $SPD$~---~single-photon detector, $PC$~---~computer. The regions of the polarization controllers are outlined in the scheme by frames $PC_{1,2}$.}
     \label{Fig: tomography}
     \end{figure}

The measurement part of the setup (Fig.~\ref{Fig: tomography}) projects the prepared ququart state onto basis states of 5 mutually unbiased bases.
Each measurement corresponds to a projector of the form:
\begin{equation}
\hat{P}_{ml}^{n} = |\psi_{nm}\rangle\langle\psi_{nm}| - |\psi_{nl}\rangle\langle\psi_{nl}|,
\label{eq:projector}
\end{equation}
where $n$ is the basis index, and $m$ and $l$ are the indices of the states within this basis. The projector given by Eq.~\eqref{eq:projector} enables deterministic discrimination between a pair of states $(\ket{\psi_{nm}}, \ket{\psi_{nl}})$. When the projection of a state onto itself is measured, the differential signal between the two detection channels takes the value $I = \pm 1$, whereas it vanishes for all other states. This approach reduces the problem of full ququart tomography to a set of 15 independent measurements.

The measurement procedure begins with transforming the input vector vortex into a state with scalar polarization and a fixed spatial structure. This transformation is performed by a polarization controller $\mathrm{PC}_1$, consisting of a quarter-wave plate $\mathrm{QWP}_5$, a half-wave plate $\mathrm{HWP}_5$, and a polarizing beam splitter $\mathrm{PBS}$. The orientation angles of the wave plates are chosen so that the initial ququart state, encoded in the four-dimensional ''polarization--phase'' space, is efficiently separated by polarization. A horizontally polarized component is formed in the upper arm after the PBS, and a vertically polarized component is formed in the lower arm. Importantly, this interferometric separation scheme preserves information about the relative phase between the two polarization qubits, which is subsequently converted into a phase difference between the detection channels.

\begin{figure}[htbp]
     \centering
     \includegraphics[width=0.55\columnwidth]{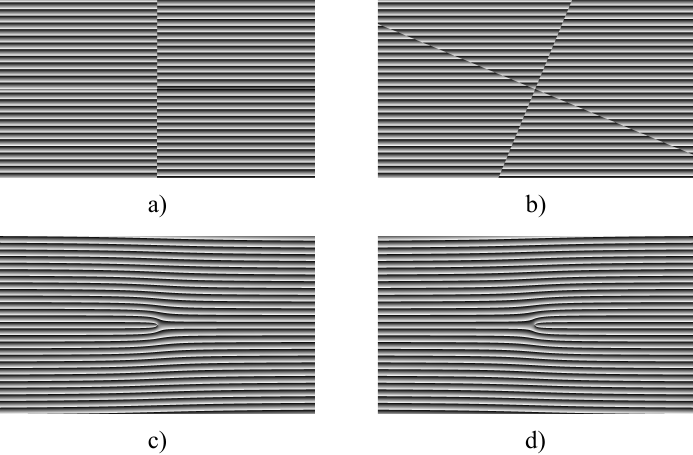}
     \caption{Examples of two-dimensional phase distributions displayed on the SLM: a) for compensating the $HG_{1,1}$ mode at an angle of $0^\circ$, b) for compensating the $HG_{1,1}$ mode at an angle of $22.5^\circ$, c) for compensating the $LG_{0,2}$ mode at an angle of $0^\circ$, d) for compensating the $LG_{0,-2}$ mode at an angle of $0^\circ$.}
     \label{Fig: phase masks}
     \end{figure}

The spatial distribution of intensity and phase in each arm after $PC_1$ is described by the Laguerre-Gaussian mode $LG_{0,2}$ for the basis IV states and by the Hermite-Gaussian mode $HG_{1,1}$ for all other states. The measured state is projected onto the target spatial mode using a spatial light modulator (SLM). An additive mixture of a blazed diffraction grating and a phase mask (Fig.~\ref{Fig: phase masks}) is displayed on the SLM, and this mixture compensates for the relative phase between the ''lobes'' of the spatial mode. The mask is designed so that all wavefront points of the target mode acquire the same phase after reflection from the SLM. Subsequent projection onto the fundamental mode of a single-mode fiber yields a maximum signal. To distinguish states from bases I, II, III, and V, the mask (Fig. ~\ref{Fig: phase masks}a,~\ref{Fig: phase masks}b) comprises 4 symmetric sectors that match the intensity distribution of the measured mode, whereas the compensating phase is applied to pairs of opposite ''lobes''. This design enables a single set of masks to distinguish all states both within one basis and across bases. For basis IV states, the SLM displays a mask (Fig.~\ref{Fig: phase masks}c,~\ref{Fig: phase masks}d) that compensates for the continuous phase distribution of the vortex structure of the $LG_{0,2}$ or $LG_{0,-2}$ mode.

Quarter-wave plates $QWP_{6,7}$ oriented at $45^\circ$ and polarizing beam splitters $PBS$ direct the radiation deterministically onto the SLM and then into the lens system. The beam then passes through a second polarization controller $PC_2$, which consists of half-wave plates $HWP_{6,7}$ in the input channels of the PBS and half-wave plates $HWP_{8,9}$ in its output channels. The controller $PC_2$ operates in two regimes. In the ''mirror'' regime, the plate $HWP_{6}$ is oriented at $0^{\circ}$ and the plate $HWP_{7}$ at $45^{\circ}$, so that horizontally polarized radiation is converted into vertically polarized radiation and reflected from the PBS. The plates $HWP_{8}$ and $HWP_{9}$ are oriented at $90^{\circ}$ and change neither the polarization nor the phase. In this regime, each beam reflected from the SLM is directed into its own single-mode fiber, and the relative phase between the inputs of the fiber beam splitter equals 0 or $\pi$, depending on which state of the measured pair the projected mode belongs to. In the controllable beam splitter regime, the plates $HWP_{6,7}$ are oriented at $22.5^{\circ}$ and convert linear polarization into the balanced superposition $1/\sqrt2 (\ket{H}+\ket{V})$, so that the radiation is split equally between the two outputs of the PBS. The plate $HWP_{8}$ is set at $0^{\circ}$ and leaves the horizontally polarized state unchanged, whereas the vertically polarized state acquires a relative phase of $\pi$. The plate $HWP_{9}$ is oriented at $45^{\circ}$ and changes the polarization to the orthogonal one. As a result, radiation with the same polarization is formed at the two outputs, but with a relative phase of 0 if it arrives at the PBS from the first input and $\pi$ if it arrives from the second. Thus, the two regimes implement two different types of encoding: phase encoding in the ''mirror'' regime and ''which path'' encoding in the controllable beam splitter regime, which in our case also converts into a phase shift. The full measurement protocol is given in Table~\ref{tab: tomo proto}.

\begin{table}[htb]
\caption{Measurement protocol for all 15 projectors. The column ''Projector'' contains the defining relations $\hat{P}_{ml}^{n}=|\psi_{nm}\rangle\langle\psi_{nm}| - |\psi_{nl}\rangle\langle\psi_{nl}|$ according to expression Eq.~\eqref{eq:projector}; the columns $\mathrm{QWP}_5$ and $\mathrm{HWP}_5$ specify the orientation angles of the corresponding plates (in degrees); the columns $\phi_H$ and $\phi_V$ contain the rotation angles of the phase masks applied to the SLM in the horizontal $\text{SLM}_H$ and vertical $\text{SLM}_H$ channels. In the basis IV, the designation ''CW'' corresponds to the mask for compensating the $LG_{0,2}$ mode, and ''CCW''~---~corresponds to the mask for compensating $LG_{0,-2}$. In all other bases, a mask for compensating the $HG_{1,1}$ mode is displayed; the column ''$PC_2$'' indicates the operating regime of the polarization controller; the columns $D_1$ and $D_2$ contain the designations of the states detected by the detectors SPD$_1$ and SPD$_2$.}
\begin{tabular}{|c|c|c|c|c|c|c|c|c|}
\hline
Basis                & Projector        & $\mathrm{QWP}_5$, $^\circ$ & $\mathrm{HWP}_5$, $^\circ$ & $\phi_H$, $^\circ$ & $\phi_V$, $^\circ$ & $PC_2$         & $D_1$             & $D_2$             \\ \hline
\multirow{3}{*}{I}   & $\hat{P}_{12}^1$ & 0                          & 0                          & 45                 & 0                  & ''mirror''     & $\ket{\psi_{11}}$ & $\ket{\psi_{12}}$ \\ \cline{2-9} 
                     & $\hat{P}_{34}^1$ & 0                          & 0                          & 90                 & 45                 & ''mirror''     & $\ket{\psi_{13}}$ & $\ket{\psi_{14}}$ \\ \cline{2-9} 
                     & $\hat{P}_{14}^1$ & 45                         & 22.5                       & 22.5               & $-22.5$            & ''mirror''     & $\ket{\psi_{11}}$ & $\ket{\psi_{14}}$ \\ \hline
\multirow{3}{*}{II}  & $\hat{P}_{13}^2$ & 0                          & 0                          & 45                 & $-45$              & ''mirror''     & $\ket{\psi_{21}}$ & $\ket{\psi_{23}}$ \\ \cline{2-9} 
                     & $\hat{P}_{24}^2$ & 0                          & 0                          & 0                  & 0                  & ''mirror''     & $\ket{\psi_{22}}$ & $\ket{\psi_{24}}$ \\ \cline{2-9} 
                     & $\hat{P}_{12}^2$ & 45                         & 22.5                       & 45                 & 0                  & ''which path'' & $\ket{\psi_{21}}$ & $\ket{\psi_{22}}$ \\ \hline
\multirow{3}{*}{III} & $\hat{P}_{14}^3$ & 0                          & 0                          & 66.5               & $-22.5$            & ''mirror''     & $\ket{\psi_{31}}$ & $\ket{\psi_{34}}$ \\ \cline{2-9} 
                     & $\hat{P}_{23}^3$ & 0                          & 0                          & 22.5               & $-66.5$            & ''mirror''     & $\ket{\psi_{32}}$ & $\ket{\psi_{33}}$ \\ \cline{2-9} 
                     & $\hat{P}_{13}^3$ & 45                         & 0                          & 22.5               & $-22.5$            & ''which path'' & $\ket{\psi_{31}}$ & $\ket{\psi_{33}}$ \\ \hline
\multirow{3}{*}{IV}  & $\hat{P}_{12}^4$ & 0                          & 0                          & $45$, CW           & $-45$, CCW         & ''which path'' & $\ket{\psi_{41}}$ & $\ket{\psi_{42}}$ \\ \cline{2-9} 
                     & $\hat{P}_{34}^4$ & 0                          & 0                          & $45$, CW           & $-45$, CW          & ''which path'' & $\ket{\psi_{43}}$ & $\ket{\psi_{44}}$ \\ \cline{2-9} 
                     & $\hat{P}_{14}^4$ & 45                         & 0                          & $66.5$, CW         & $22.5$, CW         & ''mirror''     & $\ket{\psi_{41}}$ & $\ket{\psi_{44}}$ \\ \hline
\multirow{3}{*}{V}   & $\hat{P}_{13}^5$ & 0                          & 0                          & 66.5               & 22.5               & ''mirror''     & $\ket{\psi_{51}}$ & $\ket{\psi_{53}}$ \\ \cline{2-9} 
                     & $\hat{P}_{24}^5$ & 0                          & 0                          & 22.5               & $-22.5$            & ''mirror''     & $\ket{\psi_{52}}$ & $\ket{\psi_{54}}$ \\ \cline{2-9} 
                     & $\hat{P}_{14}^5$ & 45                         & 0                          & 45                 & 0                  & ''mirror''     & $\ket{\psi_{51}}$ & $\ket{\psi_{54}}$ \\ \hline
\end{tabular}
\label{tab: tomo proto}
\end{table}

An optical scheme consisting of lenses $L_3, L_4, L_5$ and an iris diaphragm is used to form the image of the target spatial mode at the end face of a single-mode fiber. A similar system couples light into the fiber in the second channel. The SLM is illuminated by a collimated beam. An additive mixture of a phase mask and a blazed diffraction grating is displayed on the modulator. The phase mask compensates for the spatial structure of the $HG_{1,1}$ or $LG_{0,\pm2}$ mode, and the period and orientation of the grating are chosen so that the radiation modulated by the SLM is focused at a single point $P$ shifted from the optical axis. The lens $L_3$ with focal length $F_3=200\ \text{mm}$ is positioned so that the Fourier transform of the field at the SLM is formed in its back focal plane. For the wavelength $\lambda=808\ \text{nm}$ and the shift $l_0=5\ \text{mm}$, the grating period is $a=\lambda F_3/l_0\approx 32.4\ \mu\text{m}$, which corresponds to about 4 SLM pixels. An aperture with a diameter of $\sim 250\ \mu\text{m}$, placed in the plane $P$, transmits the first diffraction order and blocks the zeroth order.

A lens system of $L_4$ (focal length $F_4=200\ \text{mm}$) and $L_5$ (focal length $F_5=10\ \text{mm}$), spaced at a relative distance of $300\ \text{mm}$, transfers the plane $P$ onto the end face of a single-mode fiber. The fiber end face lies in the back focal plane of $L_5$ at a distance of 10~mm from the lens. This system forms an image of the plane $P$ with transverse magnification $M=F_5/F_4=0.05$. The beam converging onto the fiber is shifted from the optical axis of the system by a distance $l_{fiber}=F_5\cdot l_0/F_4=250\ \mu\text{m}$. A two-axis collimator compensates for this shift and allows the fiber face to be translated in a plane orthogonal to the optical axis. The numerical aperture of the converging beam is $NA_{fiber}=d_0/(2F_3)\cdot 1/M=0.1$, where $d_0\sim2\ \text{mm}$ is the diameter of the spatial mode ''lobe''. The fiber used here, P1-780PM-FC, has a numerical aperture of $0.12$ and a core diameter of $\approx5\ \mu\text{m}$. The transverse mode-size mismatch reduces the coupling efficiency, but the contrast of the interference pattern remains high. Thus, the measurement section of the setup is essentially a Mach-Zehnder interferometer with one bulk PBS and one fiber beam splitter FBS. The outputs of the FBS are connected to two avalanche photodiodes operating in the single-photon counting mode ($SPD_1$, $SPD_2$), and the signals from the detectors are fed to the coincidence circuit (correlation electronics).

The average values of the projectors given by Eq.~\eqref{eq:projector} are thus proportional to the normalized difference of the photon counts of the two detectors as follows:
\begin{equation}
I = \frac{N_{\text{SPD}_1} - N_{\text{SPD}_2}}{N_{\text{SPD}_1} + N_{\text{SPD}_2}},
\label{eq:metric}
\end{equation}
where $N_{SPD_{1}}$ and $N_{SPD_2}$ are the photon counts of the first and second detectors, respectively. When the state is projected onto itself, either constructive interference (counts only in $SPD_1$) or destructive interference (counts only in $SPD_2$) occurs, and the difference of the photon counts takes the value $I=\pm1$. For all other states, the photon counts after the fiber beam splitter are distributed equally between the detectors, yelding $I=0$.

Since the measurement procedure for a particular projector depends on the relative phase in the first interferometer (Fig.~\ref{Fig: tomography}), the optical scheme is calibrated first. For this purpose, the state $\ket{\Psi_{11}}$ is prepared in the generation scheme. The measurement scheme is then tuned to detect this state. The plates $QWP_5, HWP_5$ are set, and the required phase masks are displayed on the SLM. The masks are adjusted to provide maximum visibility of the interference pattern at the fiber end face, and this visibility manifests itself in the total photon count from both detectors. The final adjustment of the relative phase is performed using the liquid-crystal cells LCC1513-B in the generation and detection schemes, together with phase correction on the SLM. Correct tuning is indicated by the maximum photon-count difference between the two detectors. It is important to note that the phase adjustment in the tomography scheme is performed once for the state $\ket{\Psi_{11}}$ and is not required for further measurements, provided that the relative phase between the interferometer arms does not drift.

\subsection{Reconstruction of the density matrix of a ququart state}

\label{sec:reconstruction}
The density matrix of a ququart state can be parameterized by 15 real numbers $s_{i,j}$, which are the coefficients of the expansion of the density matrix in the basis of tensor products of Pauli matrices:
\begin{equation}
\hat{\rho} = \frac{1}{4} \sum_{i,j=1}^{4} s_{i,j}\,\sigma_i \otimes \sigma_j,
\label{eq:rho-param}
\end{equation}
where $\sigma_1 = \sigma_x$, $\sigma_2 = \sigma_y$, $\sigma_3 = \sigma_z$, $\sigma_4 = \mathds{1}$, and the coefficient $s_{4,4} = 1$ is fixed by the trace normalization condition $\mathrm{Tr}\,\hat{\rho} = 1$.

Each of the 15 measurements described in Section~\ref{sec:measurement} corresponds to the expectation value of the projector $\hat{P}_{ml}^{n} = |\psi_{nm}\rangle\langle\psi_{nm}| - |\psi_{nl}\rangle\langle\psi_{nl}|$, which is linear in the coefficients $s_{i,j}$:
\begin{align}
\label{eq:intensityMeasurements}
I_{12}^1=\langle \hat{P}_{12}^1\rangle &= \tfrac{1}{2}(s_{1,2} - s_{2,3}), &
I_{34}^1=\langle \hat{P}_{34}^1\rangle &= \tfrac{1}{2}(-s_{1,2} - s_{2,3}), &
I_{14}^1=\langle \hat{P}_{14}^1\rangle &= \tfrac{1}{2}(-s_{2,3} + s_{3,1}), \nonumber \\
I_{13}^2=\langle \hat{P}_{13}^2\rangle &= \tfrac{1}{2}(s_{1,1} + s_{1,4}), &
I_{24}^2=\langle \hat{P}_{24}^2\rangle &= \tfrac{1}{2}(s_{1,1} - s_{1,4}), &
I_{12}^2=\langle \hat{P}_{12}^2\rangle &= \tfrac{1}{2}(s_{1,4} + s_{4,1}), \nonumber \\
I_{14}^3=\langle \hat{P}_{14}^3\rangle &= \tfrac{1}{2}(-s_{2,2} + s_{2,4}), &
I_{23}^3=\langle \hat{P}_{23}^3\rangle &= \tfrac{1}{2}(s_{2,2} + s_{2,4}), &
I_{13}^3=\langle \hat{P}_{13}^3\rangle &= \tfrac{1}{2}(s_{2,4} - s_{4,2}), \\
I_{12}^4=\langle \hat{P}_{12}^4\rangle &= \tfrac{1}{2}(s_{3,3} + s_{3,4}), &
I_{34}^4=\langle \hat{P}_{34}^4\rangle &= \tfrac{1}{2}(-s_{3,3} + s_{3,4}), &
I_{14}^4=\langle \hat{P}_{14}^4\rangle &= \tfrac{1}{2}(s_{3,4} + s_{4,3}), \nonumber \\
I_{13}^5=\langle \hat{P}_{13}^5\rangle &= \tfrac{1}{2}(-s_{1,3} - s_{2,1}), &
I_{24}^5=\langle \hat{P}_{24}^5\rangle &= \tfrac{1}{2}(-s_{1,3} + s_{2,1}), &
I_{14}^5=\langle \hat{P}_{14}^5\rangle &= \tfrac{1}{2}(-s_{1,3} - s_{3,2}). \nonumber
\end{align}
The projectors given in Eq.~\eqref{eq:intensityMeasurements} correspond to procedures for discriminating specific pairs of basis states in each of the 5 MUB in the proposed experimental scheme. We note that pairs of states being discriminated differ from one basis to another.

The experimentally measured quantities $I_{ml}^n$, defined for each chosen pair of vectors in any MUB according to Eq.~\eqref{eq:metric}, are estimates of the expectation values of the corresponding projectors and form a vector $\mathbf{I} = (I_{11}^1, \ldots, I_{14}^5)^{\mathrm{T}}$ of dimension 15. The relations between $\mathbf{I}$ and the vector of unknown parameters $\mathbf{s} = (s_{1,1}, s_{1,2}, s_{1,3}, s_{1,4}, s_{2,1}, \ldots, s_{4,3})^{\mathrm{T}}$ can be written in matrix form as follows:
\begin{equation}
\mathbf{I} = \mathbf{M} \cdot \mathbf{s},
\label{eq:linear-system}
\end{equation}
where $\mathbf{M}$ is a real $15 \times 15$ matrix whose elements are determined by the right-hand sides of the relations above. The matrix $\mathbf{M}$ is nonsingular, which makes it possible to uniquely reconstruct the parameter vector:
\begin{equation}
\mathbf{s} = \mathbf{M}^{-1} \cdot \mathbf{I}.
\end{equation}
The obtained values of $s_{i,j}$ are substituted into Eq.~\eqref{eq:rho-param}, yielding the density matrix $\hat{\rho}_\mathrm{exp}$ reconstructed by the linear inversion method.
 
However, as shown in Ref.~\cite{Bogdanov2022}, the matrix $\hat{\rho}_\mathrm{exp}$ obtained by linear inversion generally does not satisfy physicality conditions. Due to statistical and systematic measurement errors, it can have small negative eigenvalues, and its trace can differ from unity. The subsequent algorithm depends on whether prior information about the purity of the reconstructed state is available. If the detected state is assumed to be pure, the tomography fidelity can be calculated without reconstructing the density matrix, directly from the expectation values of tensor products of Pauli operators~\cite{Flammia2011, Bantysh2020}. This approach makes it possible to estimate the fidelity of the experimentally measured state with respect to pure state, assuming that the appearance of a statistical mixture is solely a consequence of measurement error. The method is based on the linear dependence of the estimated quantity on the elements of the density matrix. Thus, linearity makes it possible to estimate the fidelity correctly despite some ''unphysicality'' of the density matrix.

However, this approach is not justified when calculating quantities that are nonlinear in $\hat{\rho}$, such as the state purity. It is also inapplicable no information is available about the purity of the target state and it must be treated as mixed. In this case, a  projection onto the set of physical states can be applied~\cite{BlumeKohout2010}. According to this procedure, the matrix is first Hermitized as follows:
\begin{equation}
\hat{\rho}_\mathrm{exp} \;\to\; \hat{\rho}_\mathrm{herm} = \tfrac{1}{2}\!\left(\hat{\rho}_\mathrm{exp} + \hat{\rho}_\mathrm{exp}^\dagger\right).
\end{equation}
Second, the matrix $\hat{\rho}_\mathrm{herm}$ is expanded in the eigenbasis, and the negative eigenvalues $\lambda_k < 0$ are replaced by zero:
\begin{equation}
\hat{\rho}_\mathrm{herm} = \sum_k \lambda_k |k\rangle\langle k| \;\to\; \hat{\rho}_{\mathrm{PSD}} = \sum_k \max(\lambda_k, 0)\,|k\rangle\langle k|.
\end{equation}
Third, the trace is normalized:
\begin{equation}
\hat{\rho}_\mathrm{norm} = \frac{\hat{\rho}_\mathrm{PSD}}{\mathrm{Tr}\,\hat{\rho}_\mathrm{PSD}}.
\label{eq:rho-norm}
\end{equation}
The resulting matrix $\hat{\rho}_\mathrm{norm}$ is positive semidefinite, Hermitian, and has unit trace, i. e., it correctly describes a physical quantum state. However, since the matrix $\rho_\mathrm{norm}$ is only the closest physical matrix to the experimentally reconstructed one, this procedure introduces a systematic bias into the calculated quantities, leading to their underestimation. Alternative approaches to the estimation of mixed states, such as the root approach~\cite{Bogdanov2022} and the least-squares method~\cite{Palmieri2024}, also introduce systematic biases in the estimated quantities.

In this paper, our interest is focused on estimating the accuracy of the tomography procedure itself for each of the 20 pure states, rather than the quality of their initial preparation. Thus, we restrict ourselves to the case of pure input states and attribute all ''imperfection'' in preparation to the imperfections in detection. We estimate the fidelity using the Uhlmann formula, which in this case reduces to the Frobenius norm:
\begin{equation}
		\mathcal{F}(\hat{\rho}_\mathrm{exp}, \hat{\rho}_\mathrm{theor}) = \left(\mathrm{Tr}\sqrt{\sqrt{\hat{\rho}_\mathrm{theor}}\,\hat{\rho}_\mathrm{exp}\sqrt{\hat{\rho}_\mathrm{theor}}}\right)^{\!2}=\mathrm{Tr}\!\left(\hat{\rho}_\mathrm{exp}\,\hat{\rho}_\mathrm{theor}\right).
		\label{eq:fidelity}
\end{equation}

Thus, we use the density matrix $\hat{\rho}_\mathrm{exp}$ obtained directly from the experiment to calculate the fidelity. As mentioned above, the state purity, in turn, is nonlinear with respect to the matrix elements of the density matrix:
\begin{equation}
		\mathcal{P}(\hat{\rho}) = \mathrm{Tr}\!\left(\hat{\rho}^2\right).
		\label{eq:purity}
\end{equation}
For the purity calculation, we therefore use the closest physical density matrix $\hat{\rho}_\mathrm{norm}$ in expression Eq.~\eqref{eq:purity}.

The errors of the measured quantities $I_{ml}^n$ are determined by the statistics of the photon counts. Since single-photon detection obeys Poisson statistics, the standard deviation of the number of counts $N$ is $\sqrt{N}$. The errors of the matrix elements of $\hat{\rho}_\mathrm{exp}$, as well as those of the purity and fidelity, are calculated by numerical error propagation~\cite{Schwemmer2015}. The reconstruction results for all 20 states, together with the purity and fidelity values and their uncertainties, are presented in Section~\ref{sec:results}.

\section{Results}

\label{sec:results}
In the experiment, observable given by Eq.~\eqref{eq:intensityMeasurements} was measured 10 times. This number of measurements is determined, on the one hand, by the signal accumulation time in each individual measurement and, on the other hand, is sufficient to analyze the statistical spread of the measured values. For each of the 20 states, the density matrix $\hat{\rho}_\mathrm{exp}$ is reconstructed by linear inversion followed by projection onto the set of physical states, as described in Section~\ref{sec:reconstruction}.
According to the procedure described in Section~\ref{sec:reconstruction}, the fidelity $\mathcal{F}$ is calculated from formula Eq.~\eqref{eq:fidelity} using the original matrix $\hat{\rho}_\mathrm{exp}$, and the purity calculated using Eq.~\eqref{eq:purity} for both the original ($\mathcal{P}_\mathrm{raw}$) and the ''corrected'' ($\mathcal{P}_\mathrm{norm}$) density matrices. The purity and fidelity values for all 20 states are listed in Table~\ref{tab:purity-fidelity}.
\LTcapwidth=0.95\textwidth
\begin{longtable}{|c|c|c|c|c|}
\caption{Purity values $\mathcal{P}$ (for the raw and ''corrected'' data) and fidelity $\mathcal{F}$ for all 20 states. The columns $\mathcal{P}_\mathrm{raw}$ and $\mathcal{F}$ contain the values obtained directly from the experimental data, and $\mathcal{P}_\mathrm{norm}$~---~contains the purity values obtained from the ''corrected'' density matrix $\hat\rho_\mathrm{norm}$.}
\label{tab:purity-fidelity}\\
\hline
State           & Basis                & $\mathcal{P}_\mathrm{raw}$ & $\mathcal{P}_\mathrm{norm}$ & $\mathcal{F}$ \\ \hline
\endfirsthead
\hline
State           & Basis              & $\mathcal{P}_\mathrm{raw}$ & $\mathcal{P}_\mathrm{norm}$ & $\mathcal{F}$ \\ \hline
\endhead
\hline
\endfoot
$|\Psi_{11}\rangle$ & \multirow{4}{*}{I}   & $0.99 \pm 0.03$            & $0.96 \pm 0.04$             & $0.992 \pm 0.013$          \\ \cline{1-1} \cline{3-5} 
$|\Psi_{12}\rangle$ &                      & $0.97 \pm 0.04$            & $0.96 \pm 0.05$             & $0.985 \pm 0.020$          \\ \cline{1-1} \cline{3-5} 
$|\Psi_{13}\rangle$ &                      & $0.98 \pm 0.06$            & $0.96 \pm 0.04$             & $0.99 \pm 0.03$            \\ \cline{1-1} \cline{3-5} 
$|\Psi_{14}\rangle$ &                      & $1.00 \pm 0.04$            & $0.95 \pm 0.03$             & $1.00 \pm 0.02$            \\ \hline
$|\Psi_{21}\rangle$ & \multirow{4}{*}{II}  & $0.98 \pm 0.02$            & $0.94 \pm 0.04$             & $0.986 \pm 0.011$          \\ \cline{1-1} \cline{3-5} 
$|\Psi_{22}\rangle$ &                      & $0.99 \pm 0.03$            & $0.96 \pm 0.03$             & $0.994 \pm 0.013$          \\ \cline{1-1} \cline{3-5} 
$|\Psi_{23}\rangle$ &                      & $0.97 \pm 0.05$            & $0.93 \pm 0.06$             & $0.99 \pm 0.02$            \\ \cline{1-1} \cline{3-5} 
$|\Psi_{24}\rangle$ &                      & $0.97 \pm 0.03$            & $0.93 \pm 0.05$             & $0.982 \pm 0.017$          \\ \hline
$|\Psi_{31}\rangle$ & \multirow{4}{*}{III} & $0.98 \pm 0.04$            & $0.96 \pm 0.05$             & $0.99 \pm 0.02$            \\ \cline{1-1} \cline{3-5} 
$|\Psi_{32}\rangle$ &                      & $1.02 \pm 0.04$            & $0.95 \pm 0.03$             & $1.008 \pm 0.020$          \\ \cline{1-1} \cline{3-5} 
$|\Psi_{33}\rangle$ &                      & $0.99 \pm 0.04$            & $0.95 \pm 0.04$             & $0.992 \pm 0.020$          \\ \cline{1-1} \cline{3-5} 
$|\Psi_{34}\rangle$ &                      & $0.99 \pm 0.03$            & $0.95 \pm 0.04$             & $0.995 \pm 0.014$          \\ \hline
$|\Psi_{41}\rangle$ & \multirow{4}{*}{IV}  & $0.98 \pm 0.02$            & $0.98 \pm 0.04$             & $0.992 \pm 0.012$          \\ \cline{1-1} \cline{3-5} 
$|\Psi_{42}\rangle$ &                      & $0.99 \pm 0.02$            & $0.97 \pm 0.04$             & $0.995 \pm 0.010$          \\ \cline{1-1} \cline{3-5} 
$|\Psi_{43}\rangle$ &                      & $0.98 \pm 0.03$            & $0.97 \pm 0.04$             & $0.988 \pm 0.017$          \\ \cline{1-1} \cline{3-5} 
$|\Psi_{44}\rangle$ &                      & $0.99 \pm 0.03$            & $0.97 \pm 0.03$             & $0.994 \pm 0.016$          \\ \hline
$|\Psi_{51}\rangle$ & \multirow{4}{*}{V}   & $0.99 \pm 0.02$            & $0.97 \pm 0.04$             & $0.995 \pm 0.012$          \\ \cline{1-1} \cline{3-5} 
$|\Psi_{52}\rangle$ &                      & $0.98 \pm 0.06$            & $0.97 \pm 0.06$             & $0.99 \pm 0.03$            \\ \cline{1-1} \cline{3-5} 
$|\Psi_{53}\rangle$ &                      & $0.99 \pm 0.06$            & $0.93 \pm 0.04$             & $0.99 \pm 0.03$            \\ \cline{1-1} \cline{3-5} 
$|\Psi_{54}\rangle$ &                      & $0.98 \pm 0.04$            & $0.93 \pm 0.05$             & $0.99 \pm 0.02$            \\ \hline
\end{longtable}
\noindent The reconstructed density matrices for all 20 states are given in the Supplementary Materials.
 
\section{Discussion}
\label{sec:discussion}

The obtained results demonstrate high and statistically uniform reconstruction accuracy for all 20 basis states of the ququart space of axially symmetric VVB. The average purity over all 20 states is $\overline{\mathcal{P}}_\mathrm{norm} \approx 0.95$, and the average fidelity is $\overline{\mathcal{F}} \approx 0.991$ (Table~\ref{tab:means}). All states are reconstructed with a purity $\mathcal{P}_\mathrm{norm} \geq 0.93$ and a fidelity $\mathcal{F} \geq 0.982$. The spread of purity values over the entire sample ($0.93$--$0.98$) and of fidelity values ($0.982$--$1.008$) is comparable to the typical statistical errors $\sigma_\mathcal{P} \sim 0.03$--$0.06$ and $\sigma_\mathcal{F} \sim 0.015$--$0.030$, which indicates the absence of statistically significant systematic differences between individual states and bases.

It should be noted that the values $\mathcal{P}_\mathrm{raw} > 1$ and $\mathcal{F} > 1$ observed for individual states (for example, $|\Psi_{32}\rangle$ with $\mathcal{P}_\mathrm{raw} = 1.02$ and $\mathcal{F} = 1.008$) have no physical sense and are an artifact of linear inversion. In the presence of nonzero statistical and systematic errors, the reconstructed matrix $\hat{\rho}_\mathrm{exp}$ does not necessarily belong to the set of physical states, and the standard metrics $\mathcal{P}$ and $\mathcal{F}$ no longer have a meaningful upper bound for it. We note that, although the $\mathcal{P}_\mathrm{raw}$ values turn out to be closer to 1, calculating the purity by this method has no rigorous justification, as we discuss in Section~\ref{sec:reconstruction}. At the same time, the overall shift of $\mathcal{P}_\mathrm{norm}$ toward lower values is a characteristic consequence of applying the procedure of projecting the density matrix onto set of physical states.

\begin{table}[htb]
\caption{Average purity values $\overline{\mathcal{P}}_\mathrm{raw}, \overline{\mathcal{P}}_\mathrm{norm}$ (for the raw and ''corrected'' data) and average fidelity $\overline{\mathcal{F}}$ for each of the 5 bases, as well as the values averaged over all 20 states.}
\begin{tabular}{|c|c|c|c|}
\hline
Basis      & $\overline{\mathcal{P}}_\mathrm{raw}$ & $\overline{\mathcal{P}}_\mathrm{norm}$ & $\overline{\mathcal{F}}$ \\ \hline
I          & $0.99 \pm 0.04$                       & $0.96 \pm 0.02$                        & $0.99 \pm 0.02$                       \\ \hline
II         & $0.98\pm 0.03$                        & $0.94 \pm 0.02$                        & $0.988 \pm 0.015$                     \\ \hline
III        & $0.99 \pm 0.04$                       & $0.95 \pm 0.02$                        & $0.996 \pm 0.019$                     \\ \hline
IV         & $0.99 \pm 0.03$                       & $0.97 \pm 0.04$                        & $0.992 \pm 0.014$                     \\ \hline
V          & $0.99 \pm 0.05$                       & $0.95 \pm 0.02$                        & $0.99 \pm 0.02$                       \\ \hline
All bases & $0.99 \pm 0.04$                       & $0.95 \pm 0.02$                        & $0.991 \pm 0.018$                     \\ \hline
\end{tabular}
\label{tab:means}
\end{table}

The averages over the bases are close to each other. For $\overline{\mathcal{P}}_\mathrm{norm}$, the averages over the bases lie in the range $0.94$--$0.97$, and for $\overline{\mathcal{F}}$, in the range $0.988$--$0.996$ (Table~\ref{tab:means}). The observed variations lie within the typical errors of the means ($\sigma_{\overline{\mathcal{P}}} \sim 0.02$, $\sigma_{\overline{\mathcal{F}}} \sim 0.015$), so the available data do not allow one basis to be identified as systematically worse or better.

Projection onto the set of physical states has a noticeable effect on the purity values. The typical shift $\Delta\mathcal{P} = \mathcal{P}_\mathrm{norm} - \mathcal{P}_\mathrm{raw}$ is $0.02$--$0.05$ and depends only weakly on the basis number. The largest shifts correspond to the states for which the original matrix $\hat{\rho}_\mathrm{exp}$ has the largest number of negative eigenvalues, as well as to the states with $\mathcal{P}_\mathrm{raw} > 1$, that is, to the cases where linear inversion gives a manifestly unphysical result. The uncertainties do not correlate with the basis number and are determined by the statistics of the photon counts in the corresponding projectors.

\begin{figure}[hbtp]
     \centering
     \includegraphics[width=0.75\columnwidth]{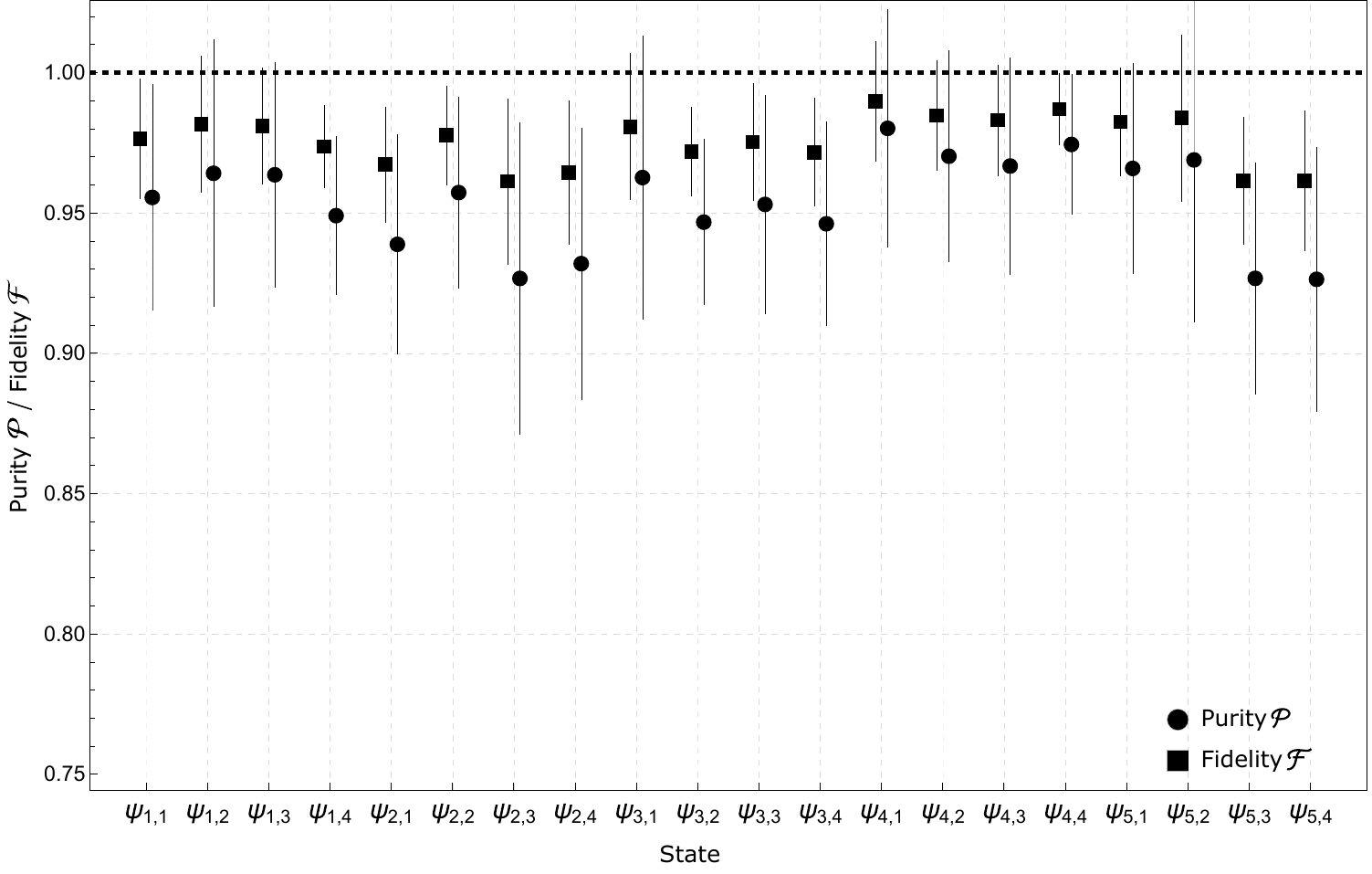}
     \caption{The purity of the ''corrected'' states $\mathcal{P}_\mathrm{norm}$ (circles) and the fidelity $\mathcal{F}$ (squares) for each of the 20 MUB
     states. The purity is shifted horizontally to the right and the fidelity to the left to separate the error bars. A horizontal dashed line is drawn at the level $\mathcal{P}_\mathrm{norm} = \mathcal{F} = 1$.}
     \label{Fig: PF comparison}
     \end{figure}

To visualize the obtained results, the purity and fidelity values after projection are presented in Fig.~\ref{Fig: PF comparison} as a scatter plot.
The diagram reveals no systematic bias for any basis. The purity and fidelity values for all 20 states form a compact group with a spread comparable to the statistical errors. Within each basis, the states show similar values in both purity and fidelity, which indicates a high degree of uniformity of the preparation and measurement procedures within a single basis.

Within each basis, purity and fidelity vary consistently. The states with the highest purity, as a rule, also have the highest fidelity, and vice versa. This indicates that the dominant contribution to the deviation from the ideal state comes from effects that simultaneously reduce purity and distort the target mode, such as the imperfection of the SLM phase masks and residual mode mismatch at the input to the single-mode fiber, rather than from phase errors, which would predominantly affect the fidelity while preserving the purity.

A pairwise comparison of the fidelity within the state pairs $(|\Psi_{nm}\rangle, |\Psi_{nl}\rangle)$ reveals no systematic excess of one fidelity over the other. The average over all ten pairs of the quantity $|\mathcal{F}^{(m)} - \mathcal{F}^{(l)}|$ does not exceed the typical statistical error $\sigma_\mathcal{F} \sim 0.02$. This indicates the absence of a systematic phase error in the measurement section of the setup, which would lead to a preferential distortion of states with a particular sign of the relative phase structure.

The observed spread in fidelity values over the entire dataset (from $0.982$ to $1.008$) most likely results not from residual differences between the states but from drift of the relative phase in the interferometers of the generation and detection schemes. In a scheme containing two Mach-Zehnder interferometers (in the preparation and measurement paths), fluctuations of the relative phase between the arms directly affect the accuracy of the projection measurements and, consequently, the fidelity value. The need for periodic calibration of the tomography scheme against the reference state $|\Psi_{11}\rangle$, noted in Section~\ref{sec:measurement}, as well as the possibility of adding an active phase compensation system to the scheme, make it possible to reduce this contribution in the future and improve the reproducibility of the results.

Thus, the experimental results indicate that the proposed MUB tomography scheme based on a polarization interferometer with spatial light modulators provides comparable characterization quality for both modes with a discrete phase structure and modes with a continuous phase structure.

\section{Conclusion}
\label{sec:conclusion}

In this paper we implement full quantum tomography of ququart states of optical vector vortices using the mutually unbiased bases technique. In contrast to previously reported experimental implementations based on q-plates, the proposed tomography scheme which is relied on a polarization Mach-Zehnder interferometer with spatial light modulators and contains no elements that fix the vortex order of the state under study. To change the vortex order, only the phase masks for the spatial light modulators need to be changed, whereas the optical scheme remains unchanged. All 20 basis states are characterized on a single setup with a single calibration of the measurement section, which ensures high reproducibility of the results and reduces systematic errors associated with drift of the setup parameters between individual measurements. We also note that measurement of physical observables with single-photon detectors requires no long signal accumulation compared with methods based on CCD-camera detection.

The generation of optical vector vortex states is performed in a modified polarization Mach-Zehnder interferometer with CCRs. Although this scheme allows all 20 MUB basis states to be generated, it has a number of drawbacks. In particular, the scheme is not universal and requires minor retuning to generate the states of basis IV. Moreover, the proposed quantum tomography scheme does not depend on the particular state generation scheme and can be configured to operate with other state-generation methods.

A comparison of the absolute purity values for the raw and ''corrected'' density matrices reveals a systematic shift. This shift is not a physical effect but is associated with the projection of the original estimate onto the set of physically admissible density matrices. It is not a drawback of the particular method used in this work; any of the currently known projection methods would also lead to a shift in the purity values. The observed spread of fidelity values over over the whole dataset most likely results from drift of the relative phase in the interferometers of the generation and detection schemes rather than from residual differences between the states. Periodic recalibration of the tomography scheme against the reference state $|\Psi_{11}\rangle$ or the addition of an active phase compensation system (for example, based on a piezoelectric actuator or a liquid-crystal cell with feedback) can reduce the phase drift and improve the reproducibility of the results. Moreover, the obtained purity and fidelity values are comparable in accuracy to data from other similar experiments~\cite{Nagali2010,Braga2026}.

Further development of the method may include active stabilization of the polarization state of the source and of the relative phase in the interferometers, improving the efficiency of mode matching between the spatial modes and the fundamental mode of the single-mode fiber, and extension of the procedure to optical vector vortex states of other orders. In particular, first-order optical vector vortices are of special interest, since such states are a common resource for encoding in quantum key distribution protocols~\cite{Reshetnikov2024,Rosales-Guzmán2018}.

\begin{acknowledgments}
This work was supported by the RSF (grant no.~26-72-10046).
\end{acknowledgments}

\bibliography{references}
\newpage

\section*{Supplementary materials. Density matrices and graphical comparison experimental results with theory}
\label{app:density-matrices}

The appendix presents all 20 reconstructed experimental density matrices together with the corresponding theoretical matrices. For each state, a visualization compares the values of the components of the experimental density matrices with those of the theoretical ones: the real parts on the left and the imaginary parts on the right. In the histograms, the outer semi-transparent blue bars correspond to theory, and the nested orange bars correspond to experiment.

\subsection*{State $\ket{\Psi_{1,1}}$}

\noindent\textbf{Experimental density matrix:}
\begin{equation*}
{\rhoexp}_{11} = \begin{pmatrix}
  0.2436 & 0.2495 - 0.01326i & 0.002405 + 0.2376i & 0.01132 - 0.2479i \\
  0.2495 + 0.01326i & 0.2764 & 0.00636 + 0.246i & 0.006284 - 0.2536i \\
  0.002405 - 0.2376i & 0.00636 - 0.246i & 0.2288 & -0.2497 - 0.00837i \\
  0.01132 + 0.2479i & 0.006284 + 0.2536i & -0.2497 + 0.00837i & 0.2513
\end{pmatrix}
\end{equation*}

\noindent\textbf{Theoretical density matrix:}
\begin{equation*}
{\rhoideal}_{11} = \begin{pmatrix}
  0.25 & 0.25 & 0.25i & -0.25i \\
  0.25 & 0.25 & 0.25i & -0.25i \\
  -0.25i & -0.25i & 0.25 & -0.25 \\
  0.25i & 0.25i & -0.25 & 0.25
\end{pmatrix}
\end{equation*}

\noindent Purity: $0.99 \pm 0.03$; fidelity: $0.992 \pm 0.013$.
\begin{figure}[h!]
  \centering
  \includegraphics[width=\textwidth]{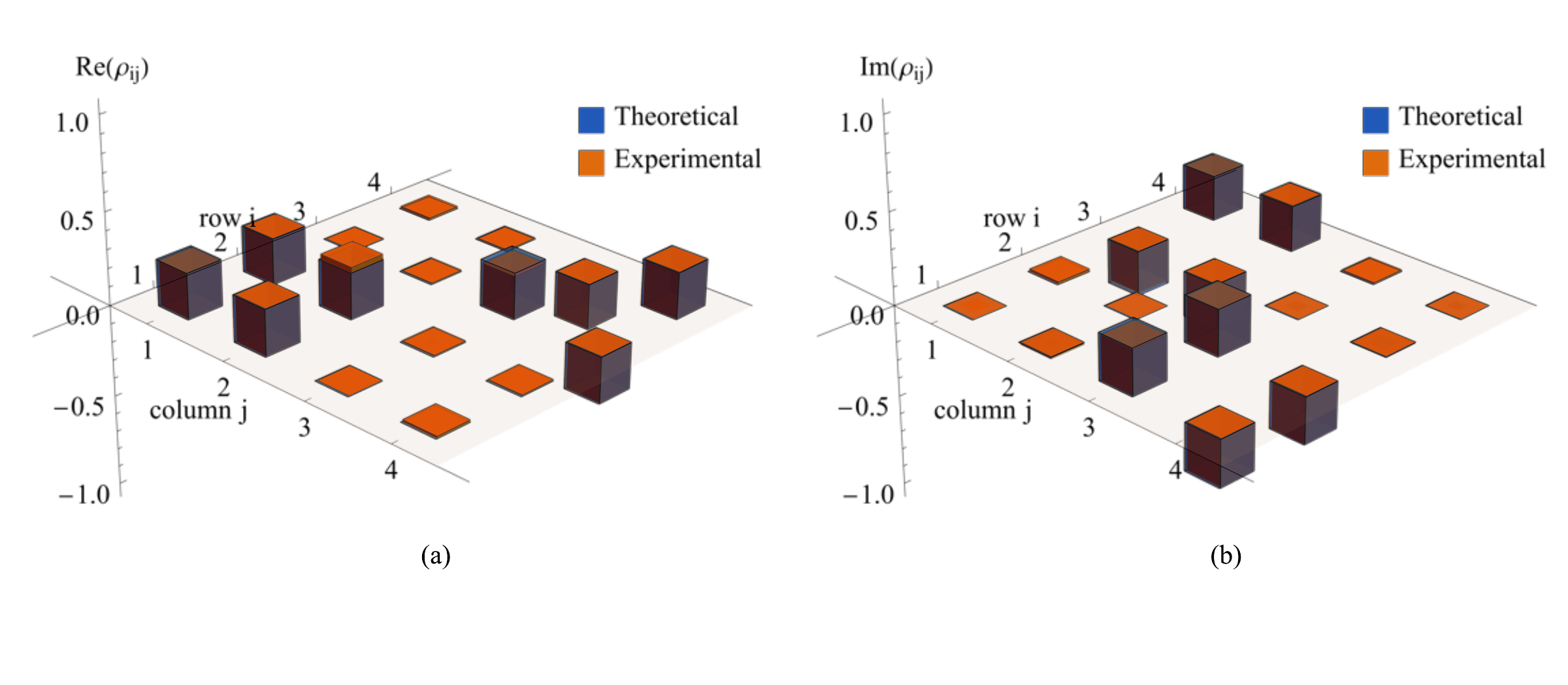}
  \caption{Real (a) and imaginary (b) parts of the density matrix for the state $\ket{\Psi_{1,1}}$. The blue (outer) bars correspond to theory, and the orange (nested) bars correspond to experiment.}
  \label{fig:A1_1}
\end{figure}

\subsection*{State $\ket{\Psi_{1,2}}$}

\noindent\textbf{Experimental density matrix:}
\begin{equation*}
{\rhoexp}_{12} = \begin{pmatrix}
  0.2543 & 0.2424 + 0.005152i & -0.00113 - 0.2453i & -0.006302 + 0.251i \\
  0.2424 - 0.005152i & 0.2527 & 0.002985 - 0.2395i & 0.003161 + 0.2497i \\
  -0.00113 + 0.2453i & 0.002985 + 0.2395i & 0.243 & -0.2414 - 0.004244i \\
  -0.006302 - 0.251i & 0.003161 - 0.2497i & -0.2414 + 0.004244i & 0.2499
\end{pmatrix}
\end{equation*}
\noindent\textbf{Theoretical density matrix:}
\begin{equation*}
{\rhoideal}_{12} = \begin{pmatrix}
  0.25 & 0.25 & -0.25i & 0.25i \\
  0.25 & 0.25 & -0.25i & 0.25i \\
  0.25i & 0.25i & 0.25 & -0.25 \\
  -0.25i & -0.25i & -0.25 & 0.25
\end{pmatrix}
\end{equation*}
\noindent Purity: $0.97 \pm 0.04$; fidelity: $0.985 \pm 0.020$.

\begin{figure}[h!]
  \centering
  \includegraphics[width=\textwidth]{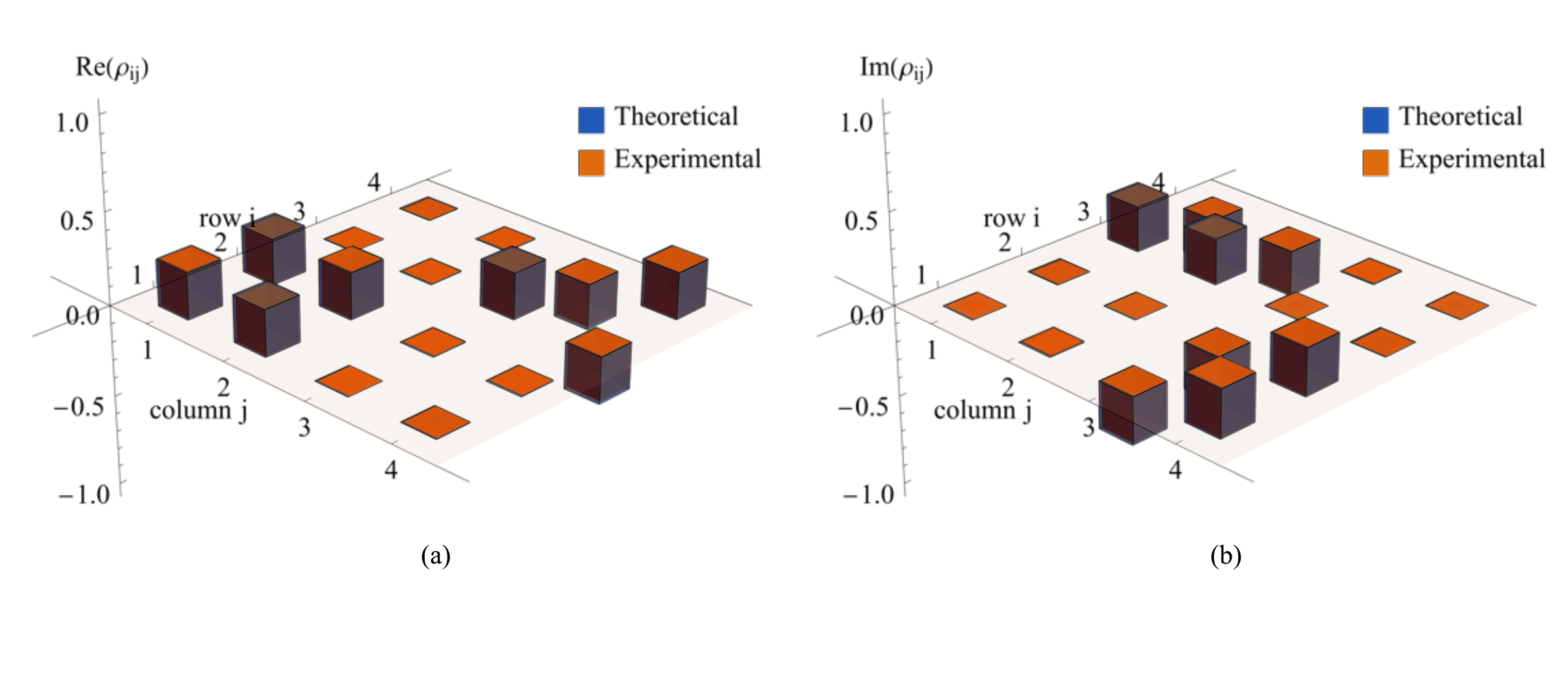}
  \caption{Real (a) and imaginary (b) parts of the density matrix for the state $\ket{\Psi_{1,2}}$. The blue (outer) bars correspond to theory, and the orange (nested) bars correspond to experiment.}
  \label{fig:A1_2}
\end{figure}

\subsection*{State $\ket{\Psi_{1,3}}$}

\noindent\textbf{Experimental density matrix:}
\begin{equation*}
{\rhoexp}_{13} = \begin{pmatrix}
  0.2437 & -0.2481 - 0.01414i & -0.01344 + 0.2451i & -0.003687 + 0.2445i \\
  -0.2481 + 0.01414i & 0.2564 & -0.01069 - 0.2465i & 0.0006562 - 0.2503i \\
  -0.01344 - 0.2451i & -0.01069 + 0.2465i & 0.2435 & 0.2482 + 0.004879i \\
  -0.003687 - 0.2445i & 0.0006562 + 0.2503i & 0.2482 - 0.004879i & 0.2564
\end{pmatrix}
\end{equation*}
\noindent\textbf{Theoretical density matrix:}
\begin{equation*}
{\rhoideal}_{13} = \begin{pmatrix}
  0.25 & -0.25 & 0.25i & 0.25i \\
  -0.25 & 0.25 & -0.25i & -0.25i \\
  -0.25i & 0.25i & 0.25 & 0.25 \\
  -0.25i & 0.25i & 0.25 & 0.25
\end{pmatrix}
\end{equation*}
\noindent Purity: $0.98 \pm 0.06$; fidelity: $0.99 \pm 0.03$.

\begin{figure}[h!]
  \centering
  \includegraphics[width=\textwidth]{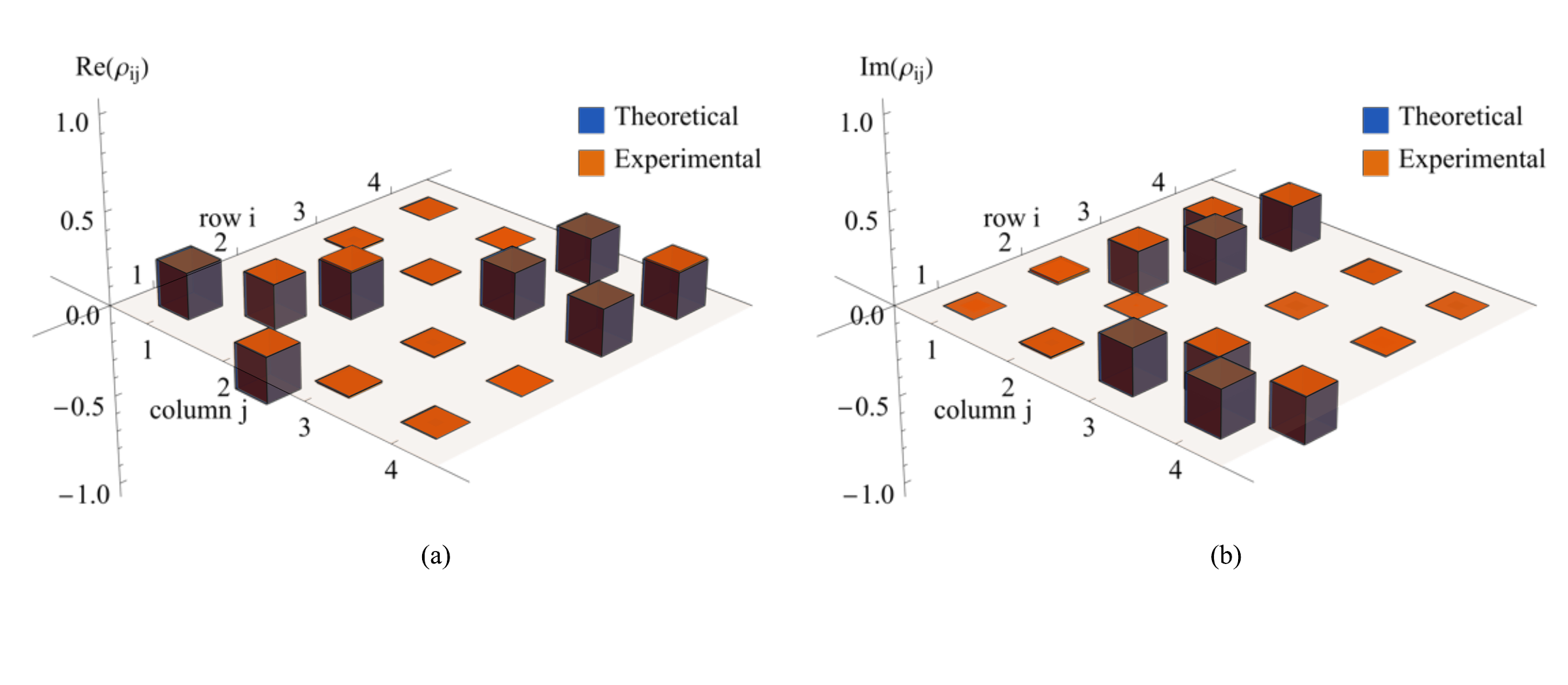}
  \caption{Real (a) and imaginary (b) parts of the density matrix for the state $\ket{\Psi_{1,3}}$. The blue (outer) bars correspond to theory, and the orange (nested) bars correspond to experiment.}
  \label{fig:A1_3}
\end{figure}

\subsection*{State $\ket{\Psi_{1,4}}$}

\noindent\textbf{Experimental density matrix:}
\begin{equation*}
{\rhoexp}_{14} = \begin{pmatrix}
  0.2358 & -0.2503 - 0.005542i & -0.002144 - 0.2316i & 0.009862 - 0.2607i \\
  -0.2503 + 0.005542i & 0.2497 & -0.006478 + 0.2543i & -0.004158 + 0.2456i \\
  -0.002144 + 0.2316i & -0.006478 - 0.2543i & 0.2644 & 0.2605 + 0.002271i \\
  0.009862 + 0.2607i & -0.004158 - 0.2456i & 0.2605 - 0.002271i & 0.2502
\end{pmatrix}
\end{equation*}
\noindent\textbf{Theoretical density matrix:}
\begin{equation*}
{\rhoideal}_{14} = \begin{pmatrix}
  0.25 & -0.25 & -0.25i & -0.25i \\
  -0.25 & 0.25 & 0.25i & 0.25i \\
  0.25i & -0.25i & 0.25 & 0.25 \\
  0.25i & -0.25i & 0.25 & 0.25
\end{pmatrix}
\end{equation*}
\noindent Purity: $1.00 \pm 0.04$; fidelity: $1.00 \pm 0.02$.

\begin{figure}[h!]
  \centering
  \includegraphics[width=\textwidth]{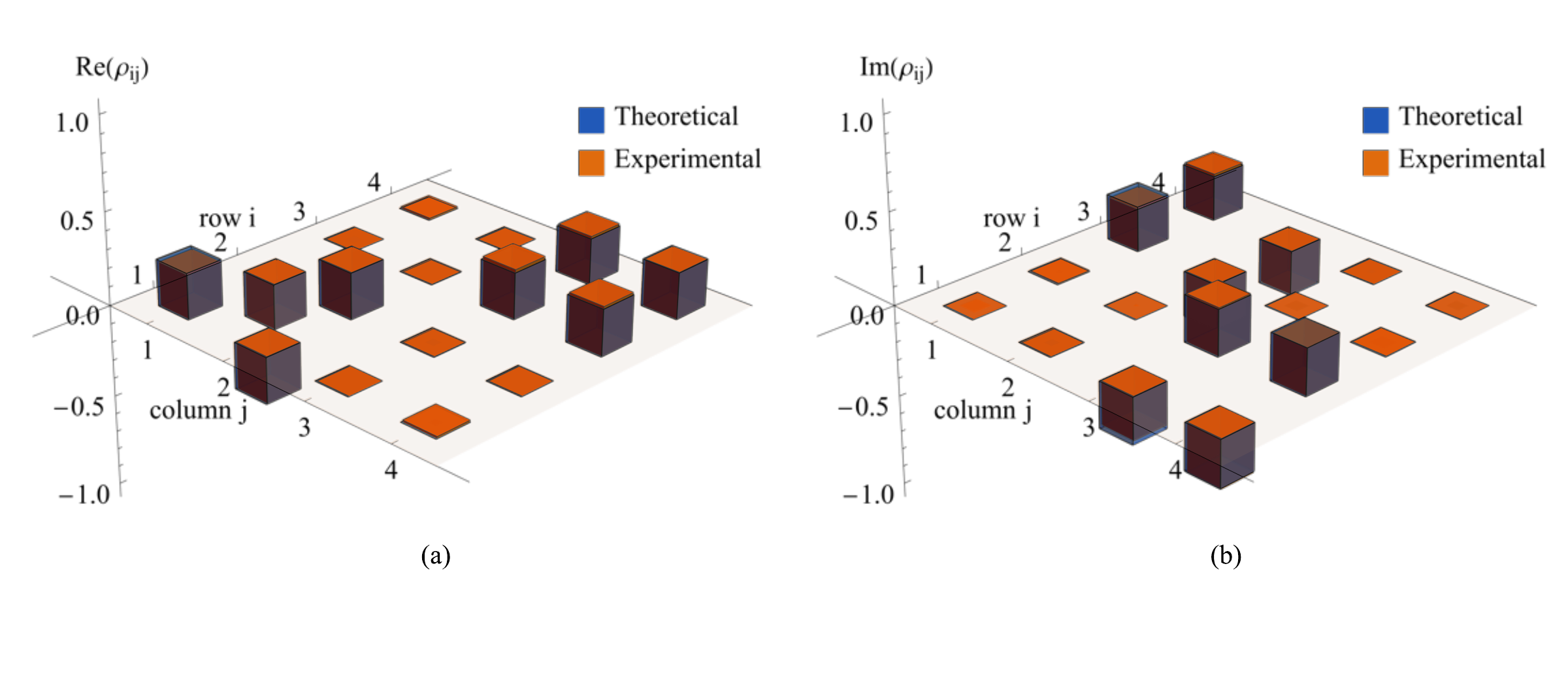}
  \caption{Real (a) and imaginary (b) parts of the density matrix for the state $\ket{\Psi_{1,4}}$. The blue (outer) bars correspond to theory, and the orange (nested) bars correspond to experiment.}
  \label{fig:A1_4}
\end{figure}

\subsection*{State $\ket{\Psi_{2,1}}$}

\noindent\textbf{Experimental density matrix:}
\begin{equation*}
{\rhoexp}_{21} = \begin{pmatrix}
  0.2617 & 0.2295 + 0.01541i & 0.2525 + 0.009285i & 0.2509 + 0.01584i \\
  0.2295 - 0.01541i & 0.2225 & 0.2367 - 0.0004318i & 0.2482 + 0.005625i \\
  0.2525 - 0.009285i & 0.2367 + 0.0004318i & 0.2716 & 0.2536 - 0.007896i \\
  0.2509 - 0.01584i & 0.2482 - 0.005625i & 0.2536 + 0.007896i & 0.2442
\end{pmatrix}
\end{equation*}
\noindent\textbf{Theoretical density matrix:}
\begin{equation*}
{\rhoideal}_{21} = \begin{pmatrix}
  0.25 & 0.25 & 0.25 & 0.25 \\
  0.25 & 0.25 & 0.25 & 0.25 \\
  0.25 & 0.25 & 0.25 & 0.25 \\
  0.25 & 0.25 & 0.25 & 0.25
\end{pmatrix}
\end{equation*}
\noindent Purity: $0.98 \pm 0.02$; fidelity: $0.986 \pm 0.011$.

\begin{figure}[h!]
  \centering
  \includegraphics[width=\textwidth]{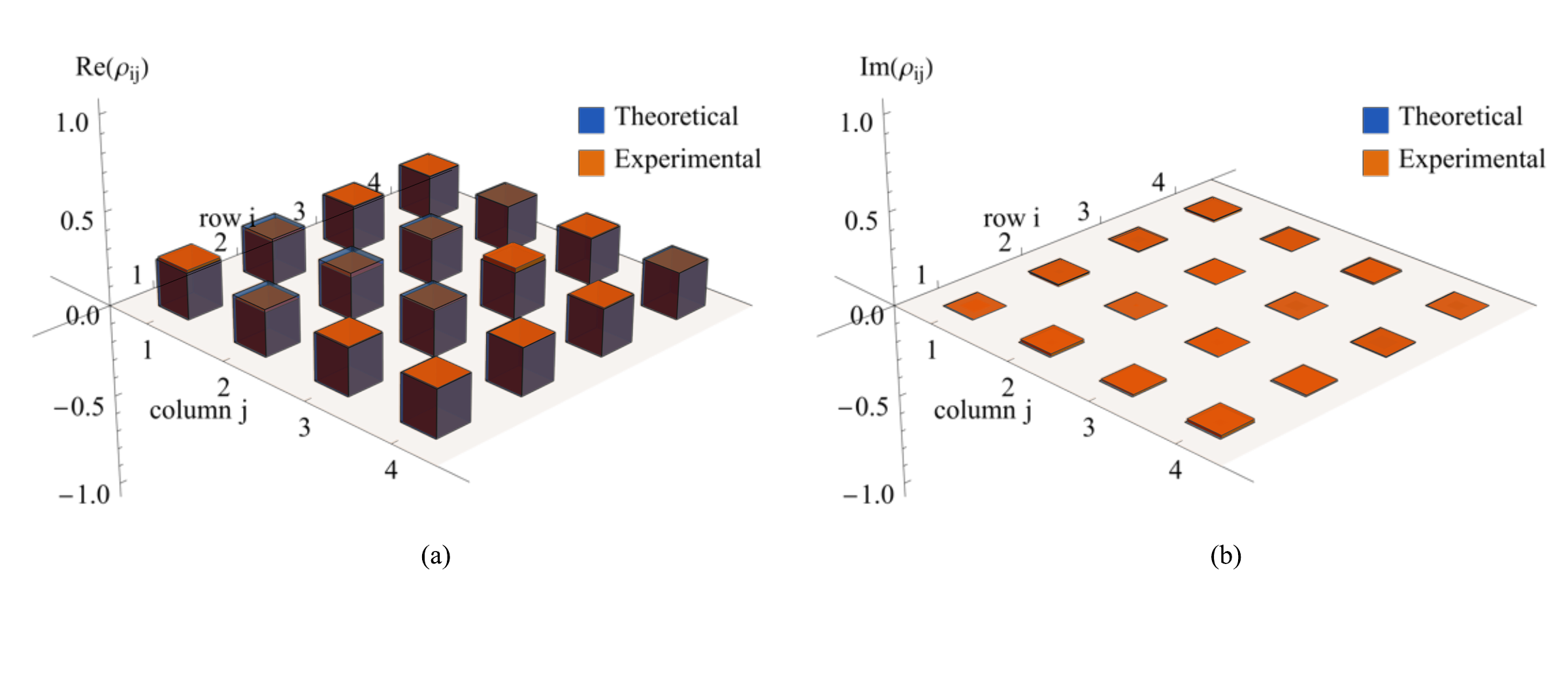}
  \caption{Real (a) and imaginary (b) parts of the density matrix for the state $\ket{\Psi_{2,1}}$. The blue (outer) bars correspond to theory, and the orange (nested) bars correspond to experiment.}
  \label{fig:A2_1}
\end{figure}

\subsection*{State $\ket{\Psi_{2,2}}$}

\noindent\textbf{Experimental density matrix:}
\begin{equation*}
{\rhoexp}_{22} = \begin{pmatrix}
  0.2501 & -0.256 + 0.001873i & -0.2485 - 0.01145i & 0.2544 + 0.007966i \\
  -0.256 - 0.001873i & 0.2699 & 0.2403 - 0.005613i & -0.2484 + 0.005862i \\
  -0.2485 + 0.01145i & 0.2403 + 0.005613i & 0.2348 & -0.2398 - 0.008058i \\
  0.2544 - 0.007966i & -0.2484 - 0.005862i & -0.2398 + 0.008058i & 0.2452
\end{pmatrix}
\end{equation*}
\noindent\textbf{Theoretical density matrix:}
\begin{equation*}
{\rhoideal}_{22} = \begin{pmatrix}
  0.25 & -0.25 & -0.25 & 0.25 \\
  -0.25 & 0.25 & 0.25 & -0.25 \\
  -0.25 & 0.25 & 0.25 & -0.25 \\
  0.25 & -0.25 & -0.25 & 0.25
\end{pmatrix}
\end{equation*}
\noindent Purity: $0.99 \pm 0.03$; fidelity: $0.994 \pm 0.013$.

\begin{figure}[h!]
  \centering
  \includegraphics[width=\textwidth]{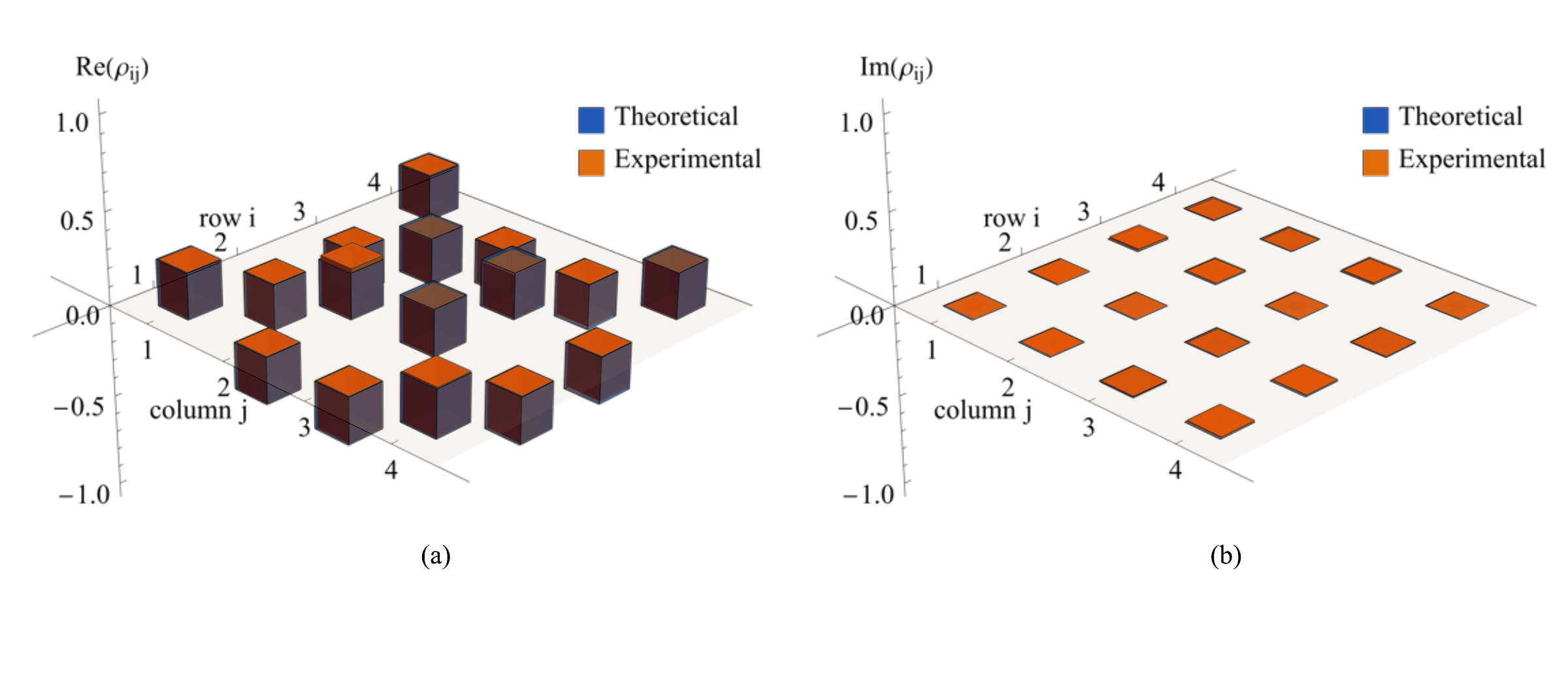}
  \caption{Real (a) and imaginary (b) parts of the density matrix for the state $\ket{\Psi_{2,2}}$. The blue (outer) bars correspond to theory, and the orange (nested) bars correspond to experiment.}
  \label{fig:A2_2}
\end{figure}

\subsection*{State $\ket{\Psi_{2,3}}$}

\noindent\textbf{Experimental density matrix:}
\begin{equation*}
{\rhoexp}_{23} = \begin{pmatrix}
  0.2354 & 0.2432 - 0.004636i & -0.2378 + 0.001163i & -0.2452 - 0.001776i \\
  0.2432 + 0.004636i & 0.2756 & -0.2571 + 0.0009426i & -0.2449 - 0.004594i \\
  -0.2378 - 0.001163i & -0.2571 - 0.0009426i & 0.234 & 0.2431 - 0.02934i \\
  -0.2452 + 0.001776i & -0.2449 + 0.004594i & 0.2431 + 0.02934i & 0.255
\end{pmatrix}
\end{equation*}
\noindent\textbf{Theoretical density matrix:}
\begin{equation*}
{\rhoideal}_{23} = \begin{pmatrix}
  0.25 & 0.25 & -0.25 & -0.25 \\
  0.25 & 0.25 & -0.25 & -0.25 \\
  -0.25 & -0.25 & 0.25 & 0.25 \\
  -0.25 & -0.25 & 0.25 & 0.25
\end{pmatrix}
\end{equation*}
\noindent Purity: $0.97 \pm 0.05$; fidelity: $0.99 \pm 0.02$.
\begin{figure}[h!]
  \centering
  \includegraphics[width=\textwidth]{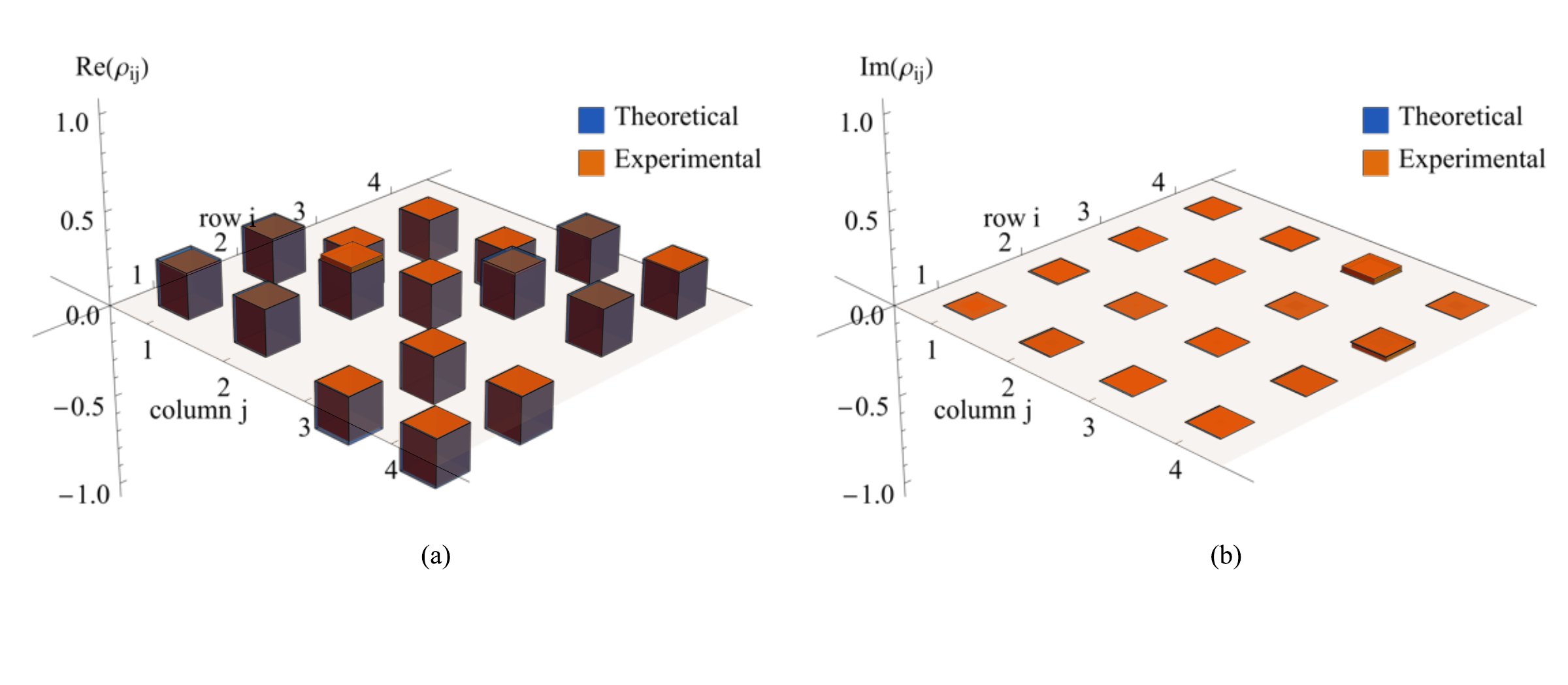}
  \caption{Real (a) and imaginary (b) parts of the density matrix for the state $\ket{\Psi_{2,3}}$. The blue (outer) bars correspond to theory, and the orange (nested) bars correspond to experiment.}
  \label{fig:A2_3}
\end{figure}

\subsection*{State $\ket{\Psi_{2,4}}$}

\noindent\textbf{Experimental density matrix:}
\begin{equation*}
{\rhoexp}_{24} = \begin{pmatrix}
  0.2446 & -0.2267 - 0.009357i & 0.2392 - 0.008035i & -0.2422 - 0.003825i \\
  -0.2267 + 0.009357i & 0.2461 & -0.2578 - 0.0129i & 0.2495 + 0.006413i \\
  0.2392 + 0.008035i & -0.2578 + 0.0129i & 0.2562 & -0.2475 - 0.007532i \\
  -0.2422 + 0.003825i & 0.2495 - 0.006413i & -0.2475 + 0.007532i & 0.2531
\end{pmatrix}
\end{equation*}
\noindent\textbf{Theoretical density matrix:}
\begin{equation*}
{\rhoideal}_{24} = \begin{pmatrix}
  0.25 & -0.25 & 0.25 & -0.25 \\
  -0.25 & 0.25 & -0.25 & 0.25 \\
  0.25 & -0.25 & 0.25 & -0.25 \\
  -0.25 & 0.25 & -0.25 & 0.25
\end{pmatrix}
\end{equation*}
\noindent Purity: $0.97 \pm 0.03$; fidelity: $0.982 \pm 0.017$.

\begin{figure}[h!]
  \centering
  \includegraphics[width=\textwidth]{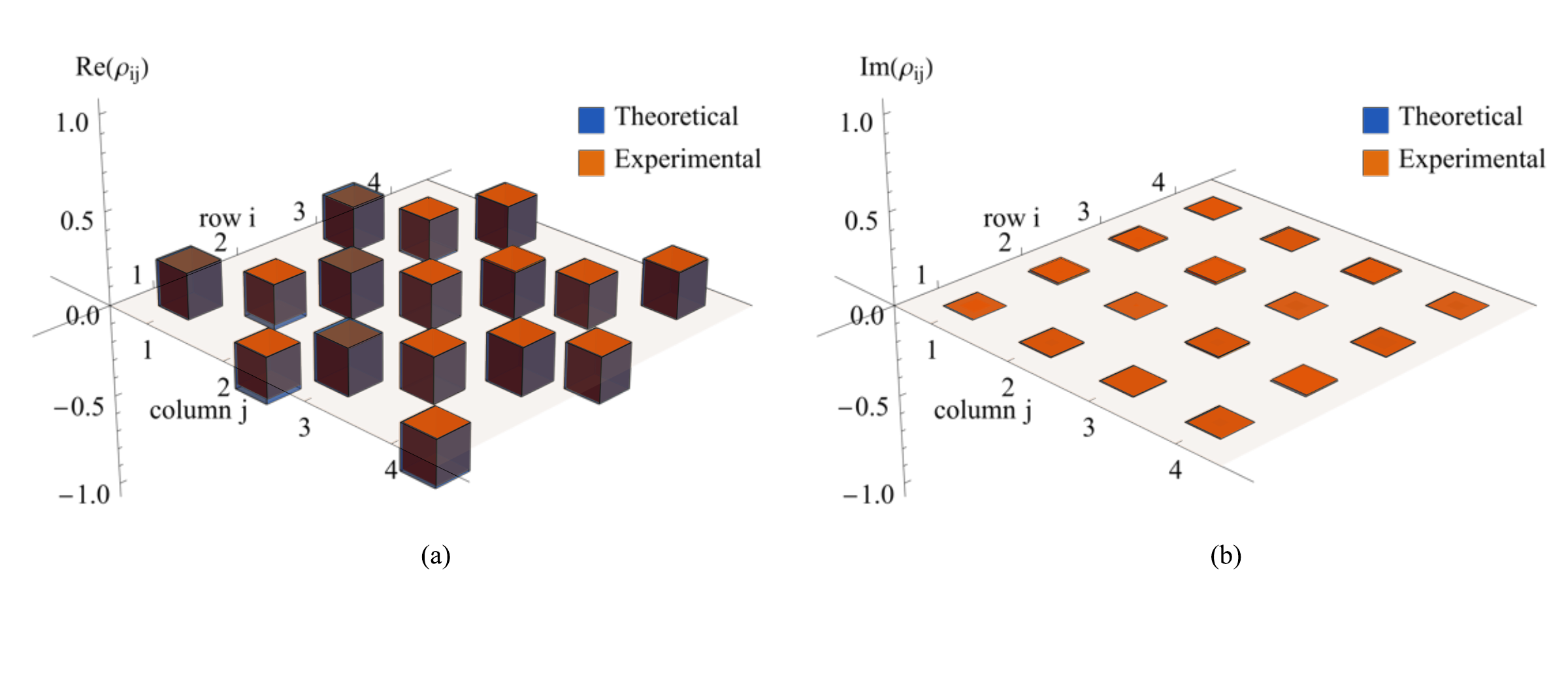}
  \caption{Real (a) and imaginary (b) parts of the density matrix for the state $\ket{\Psi_{2,4}}$. The blue (outer) bars correspond to theory, and the orange (nested) bars correspond to experiment.}
  \label{fig:A2_4}
\end{figure}

\subsection*{State $\ket{\Psi_{3,1}}$}

\noindent\textbf{Experimental density matrix:}
\begin{equation*}
{\rhoexp}_{31} = \begin{pmatrix}
  0.2507 & 0.006463 + 0.2428i & -0.006958 - 0.2531i & 0.2424 - 0.00005357i \\
  0.006463 - 0.2428i & 0.2413 & -0.248 + 0.005467i & -0.009635 - 0.2462i \\
  -0.006958 + 0.2531i & -0.248 - 0.005467i & 0.2597 & 0.008026 + 0.2493i \\
  0.2424 + 0.00005357i & -0.009635 + 0.2462i & 0.008026 - 0.2493i & 0.2482
\end{pmatrix}
\end{equation*}
\noindent\textbf{Theoretical density matrix:}
\begin{equation*}
{\rhoideal}_{31} = \begin{pmatrix}
  0.25 & 0.25i & -0.25i & 0.25 \\
  -0.25i & 0.25 & -0.25 & -0.25i \\
  0.25i & -0.25 & 0.25 & 0.25i \\
  0.25 & 0.25i & -0.25i & 0.25
\end{pmatrix}
\end{equation*}
\noindent Purity: $0.98 \pm 0.04$; fidelity: $0.99 \pm 0.02$.

\begin{figure}[h!]
  \centering
  \includegraphics[width=\textwidth]{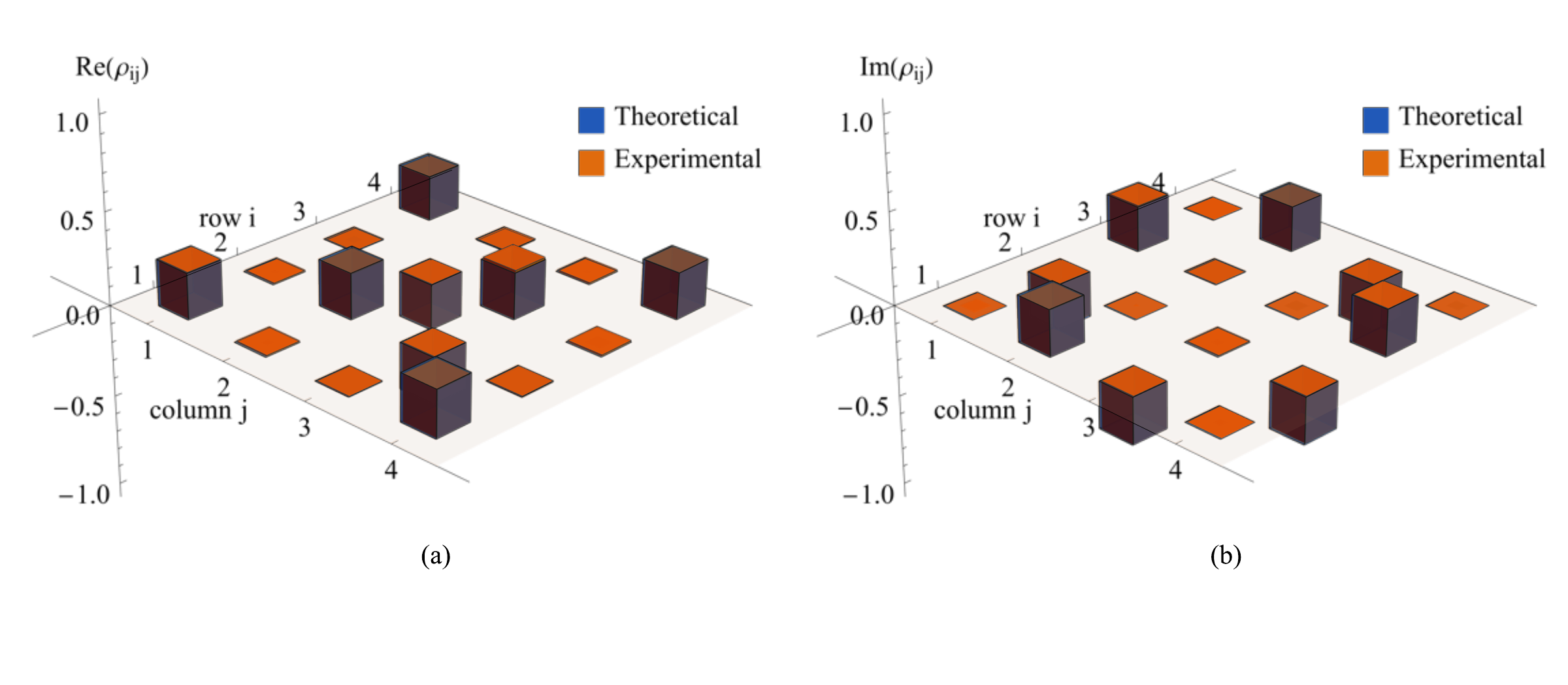}
  \caption{Real (a) and imaginary (b) parts of the density matrix for the state $\ket{\Psi_{3,1}}$. The blue (outer) bars correspond to theory, and the orange (nested) bars correspond to experiment.}
  \label{fig:A3_1}
\end{figure}

\subsection*{State $\ket{\Psi_{3,2}}$}

\noindent\textbf{Experimental density matrix:}
\begin{equation*}
{\rhoexp}_{32} = \begin{pmatrix}
  0.2489 & -0.01245 - 0.247i & -0.002629 - 0.2418i & -0.2487 + 0.005773i \\
  -0.01245 + 0.247i & 0.2548 & 0.2484 - 0.01033i & -0.006477 - 0.2548i \\
  -0.002629 + 0.2418i & 0.2484 + 0.01033i & 0.2485 & 0.02946 - 0.2754i \\
  -0.2487 - 0.005773i & -0.006477 + 0.2548i & 0.02946 + 0.2754i & 0.2478
\end{pmatrix}
\end{equation*}
\noindent\textbf{Theoretical density matrix:}
\begin{equation*}
{\rhoideal}_{32} = \begin{pmatrix}
  0.25 & -0.25i & -0.25i & -0.25 \\
  0.25i & 0.25 & 0.25 & -0.25i \\
  0.25i & 0.25 & 0.25 & -0.25i \\
  -0.25 & 0.25i & 0.25i & 0.25
\end{pmatrix}
\end{equation*}
\noindent Purity: $1.02 \pm 0.04$; fidelity: $1.008 \pm 0.020$.

\begin{figure}[h!]
  \centering
  \includegraphics[width=\textwidth]{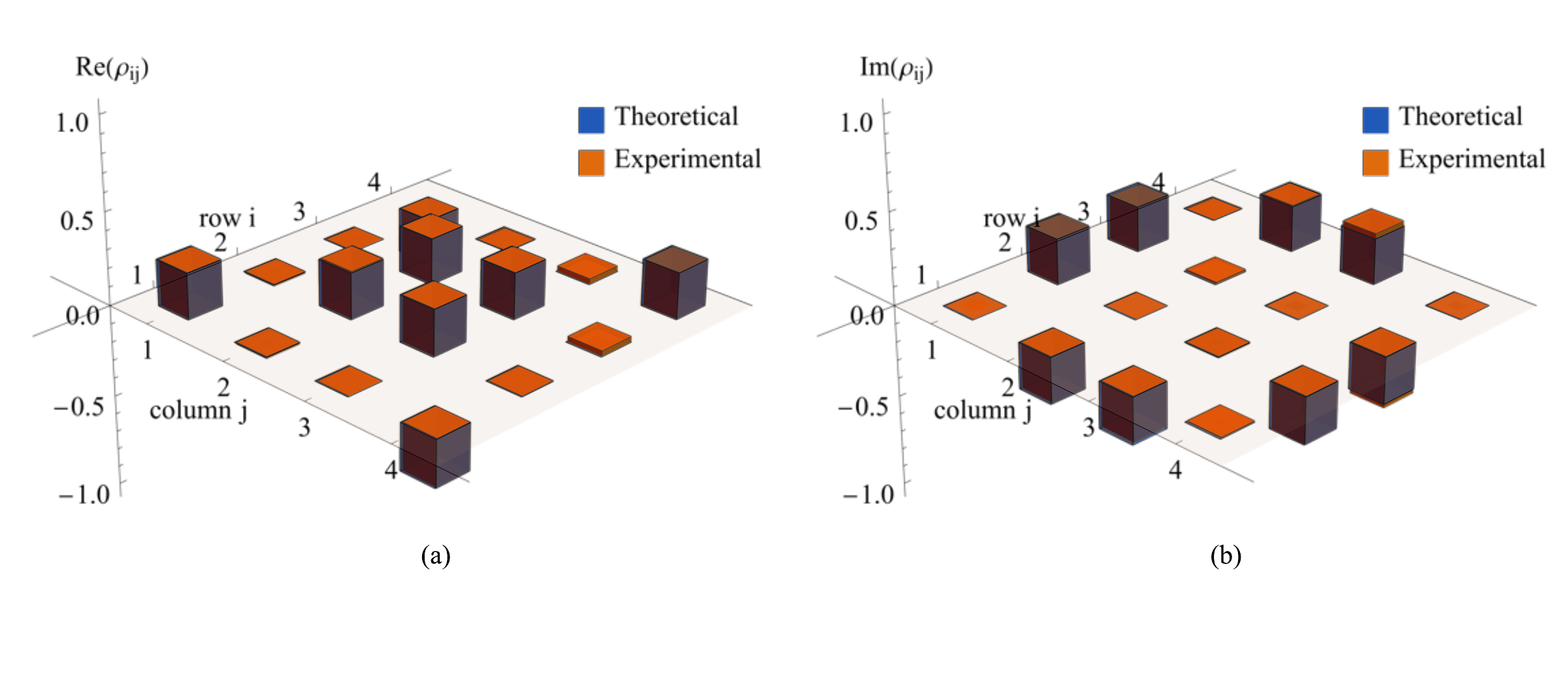}
  \caption{Real (a) and imaginary (b) parts of the density matrix for the state $\ket{\Psi_{3,2}}$. The blue (outer) bars correspond to theory, and the orange (nested) bars correspond to experiment.}
  \label{fig:A3_2}
\end{figure}

\subsection*{State $\ket{\Psi_{3,3}}$}

\noindent\textbf{Experimental density matrix:}
\begin{equation*}
{\rhoexp}_{33} = \begin{pmatrix}
  0.2581 & 0.006725 - 0.2351i & 0.01013 + 0.2521i & 0.2396 - 0.00702i \\
  0.006725 + 0.2351i & 0.239 & -0.2581 + 0.007278i & -0.007469 + 0.2436i \\
  0.01013 - 0.2521i & -0.2581 - 0.007278i & 0.2607 & 0.01264 - 0.2558i \\
  0.2396 + 0.00702i & -0.007469 - 0.2436i & 0.01264 + 0.2558i & 0.2422
\end{pmatrix}
\end{equation*}
\noindent\textbf{Theoretical density matrix:}
\begin{equation*}
{\rhoideal}_{33} = \begin{pmatrix}
  0.25 & -0.25i & 0.25i & 0.25 \\
  0.25i & 0.25 & -0.25 & 0.25i \\
  -0.25i & -0.25 & 0.25 & -0.25i \\
  0.25 & -0.25i & 0.25i & 0.25
\end{pmatrix}
\end{equation*}
\noindent Purity: $0.99 \pm 0.04$; fidelity: $0.992 \pm 0.020$.

\begin{figure}[h!]
  \centering
  \includegraphics[width=\textwidth]{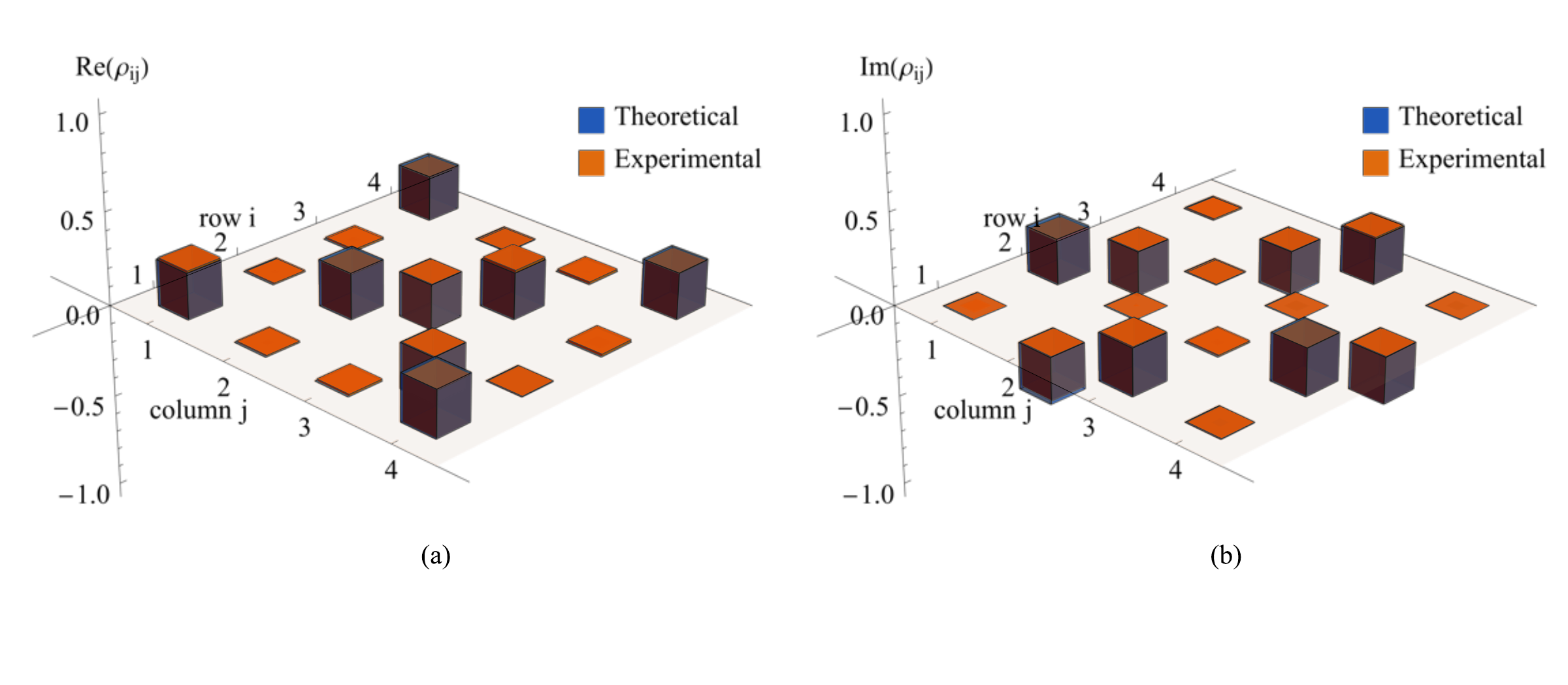}
  \caption{Real (a) and imaginary (b) parts of the density matrix for the state $\ket{\Psi_{3,1}}$. The blue (outer) bars correspond to theory, and the orange (nested) bars correspond to experiment.}
  \label{fig:A3_3}
\end{figure}

\subsection*{State $\ket{\Psi_{3,4}}$}

\noindent\textbf{Experimental density matrix:}
\begin{equation*}
{\rhoexp}_{34} = \begin{pmatrix}
  0.2481 & 0.02799 + 0.2523i & -0.006735 + 0.241i & -0.2403 - 0.000718i \\
  0.02799 - 0.2523i & 0.2571 & 0.2503 - 0.001182i & -0.01406 + 0.2553i \\
  -0.006735 - 0.241i & 0.2503 + 0.001182i & 0.2353 & 0.001453 + 0.2511i \\
  -0.2403 + 0.000718i & -0.01406 - 0.2553i & 0.001453 - 0.2511i & 0.2595
\end{pmatrix}
\end{equation*}
\noindent\textbf{Theoretical density matrix:}
\begin{equation*}
{\rhoideal}_{34} = \begin{pmatrix}
  0.25 & 0.25i & 0.25i & -0.25 \\
  -0.25i & 0.25 & 0.25 & 0.25i \\
  -0.25i & 0.25 & 0.25 & 0.25i \\
  -0.25 & -0.25i & -0.25i & 0.25
\end{pmatrix}
\end{equation*}
\noindent Purity: $0.99 \pm 0.03$; fidelity: $0.995 \pm 0.014$.

\begin{figure}[h!]
  \centering
  \includegraphics[width=\textwidth]{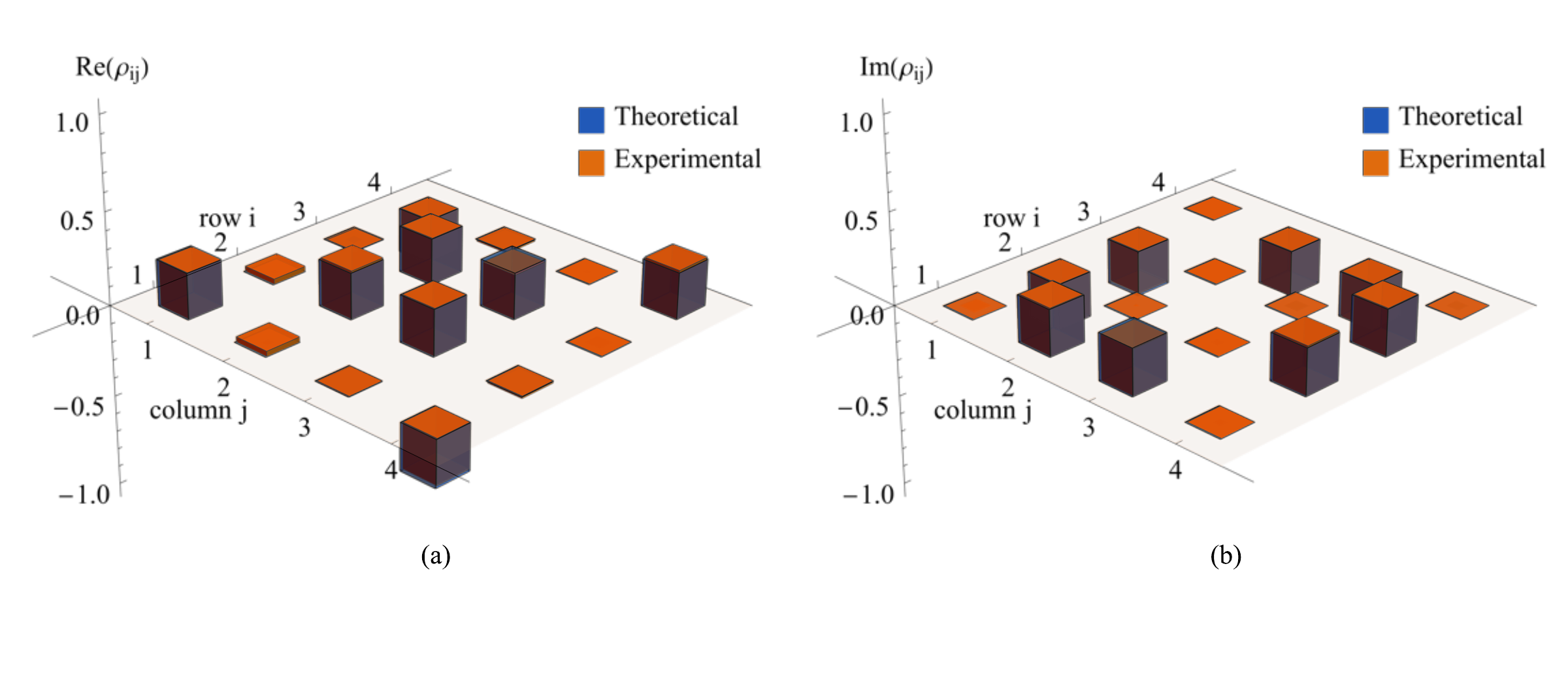}
  \caption{Real (a) and imaginary (b) parts of the density matrix for the state $\ket{\Psi_{3,4}}$. The blue (outer) bars correspond to theory, and the orange (nested) bars correspond to experiment.}
  \label{fig:A3_4}
\end{figure}

\subsection*{State $\ket{\Psi_{4,1}}$}

\noindent\textbf{Experimental density matrix:}
\begin{equation*}
{\rhoexp}_{41} = \begin{pmatrix}
  0.9915 & -0.00075 - 0.01225i & -0.005 - 0.0025i & 0.00225 - 0.00375i \\
  -0.00075 + 0.01225i & 0.0025 & -0.00175 + 0.00225i & 0.0005 - 0.0005i \\
  -0.005 + 0.0025i & -0.00175 - 0.00225i & 0.0015 & -0.00275 + 0.00125i \\
  0.00225 + 0.00375i & 0.0005 + 0.0005i & -0.00275 - 0.00125i & 0.0045
\end{pmatrix}
\end{equation*}
\noindent\textbf{Theoretical density matrix:}
\begin{equation*}
{\rhoideal}_{41} = \begin{pmatrix}
  1. & 0 & 0 & 0 \\
  0 & 0 & 0 & 0 \\
  0 & 0 & 0 & 0 \\
  0 & 0 & 0 & 0
\end{pmatrix}
\end{equation*}
\noindent Purity: $0.98 \pm 0.02$; fidelity: $0.992 \pm 0.012$.

\begin{figure}[h!]
  \centering
  \includegraphics[width=\textwidth]{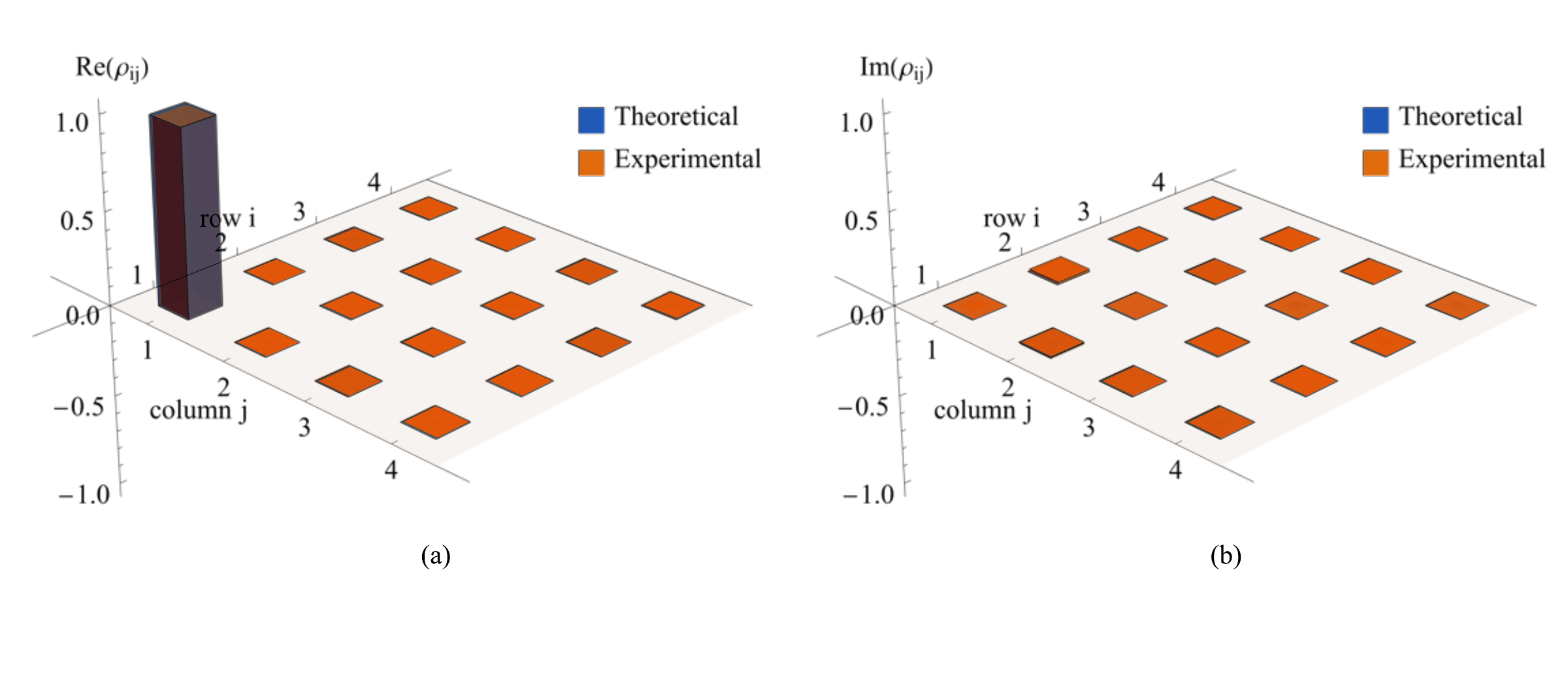}
  \caption{Real (a) and imaginary (b) parts of the density matrix for the state $\ket{\Psi_{4,1}}$. The blue (outer) bars correspond to theory, and the orange (nested) bars correspond to experiment.}
  \label{fig:A4_1}
\end{figure}

\subsection*{State $\ket{\Psi_{4,2}}$}

\noindent\textbf{Experimental density matrix:}
\begin{equation*}
{\rhoexp}_{42} = \begin{pmatrix}
  0.00425 & -0.00925 + 0.00575i & 0.00375 + 0.00425i & 0.0005 - 0.0045i \\
  -0.00925 - 0.00575i & 0.00125 & -0.0035 & 0.00225 - 0.00125i \\
  0.00375 - 0.00425i & -0.0035 & 0.9953 & 0.01325 + 0.00125i \\
  0.0005 + 0.0045i & 0.00225 + 0.00125i & 0.01325 - 0.00125i & -0.00075
\end{pmatrix}
\end{equation*}
\noindent\textbf{Theoretical density matrix:}
\begin{equation*}
{\rhoideal}_{42} = \begin{pmatrix}
  0 & 0 & 0 & 0 \\
  0 & 0 & 0 & 0 \\
  0 & 0 & 1. & 0 \\
  0 & 0 & 0 & 0
\end{pmatrix}
\end{equation*}
\noindent Purity: $0.99 \pm 0.02$; fidelity: $0.995 \pm 0.010$.

\begin{figure}[h!]
  \centering
  \includegraphics[width=\textwidth]{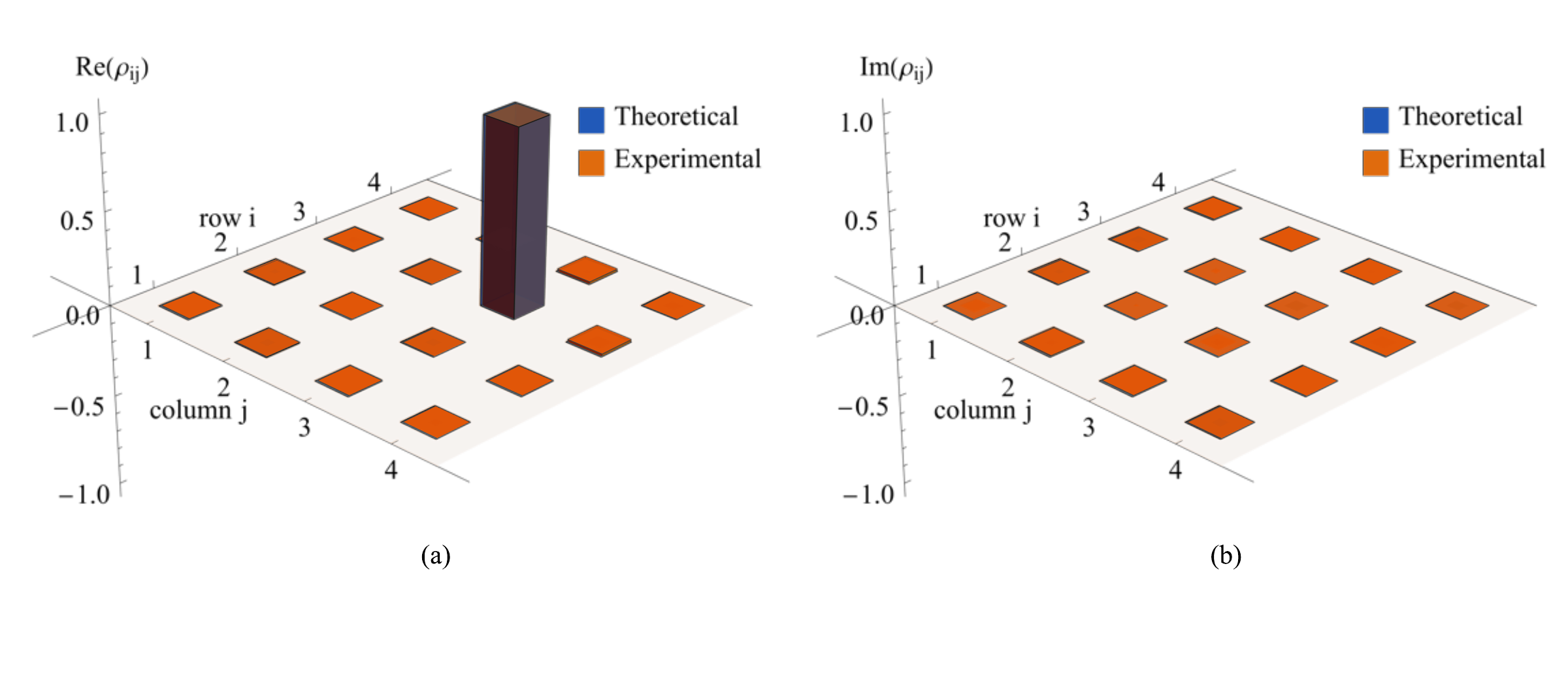}
  \caption{Real (a) and imaginary (b) parts of the density matrix for the state $\ket{\Psi_{4,2}}$. The blue (outer) bars correspond to theory, and the orange (nested) bars correspond to experiment.}
  \label{fig:A4_2}
\end{figure}

\subsection*{State $\ket{\Psi_{4,3}}$}

\noindent\textbf{Experimental density matrix:}
\begin{equation*}
{\rhoexp}_{43} = \begin{pmatrix}
  0.005 & -0.004 + 0.00375i & 0.00325 + 0.0045i & -0.001 + 0.00375i \\
  -0.004 - 0.00375i & 0.988 & -0.0045 + 0.00225i & -0.00175 - 0.003i \\
  0.00325 - 0.0045i & -0.0045 - 0.00225i & 0.009 & 0.0005 - 0.00225i \\
  -0.001 - 0.00375i & -0.00175 + 0.003i & 0.0005 + 0.00225i & -0.002
\end{pmatrix}
\end{equation*}
\noindent\textbf{Theoretical density matrix:}
\begin{equation*}
{\rhoideal}_{43} = \begin{pmatrix}
  0 & 0 & 0 & 0 \\
  0 & 1. & 0 & 0 \\
  0 & 0 & 0 & 0 \\
  0 & 0 & 0 & 0
\end{pmatrix}
\end{equation*}
\noindent Purity: $0.98 \pm 0.03$; fidelity: $0.988 \pm 0.017$.

\begin{figure}[h!]
  \centering
  \includegraphics[width=\textwidth]{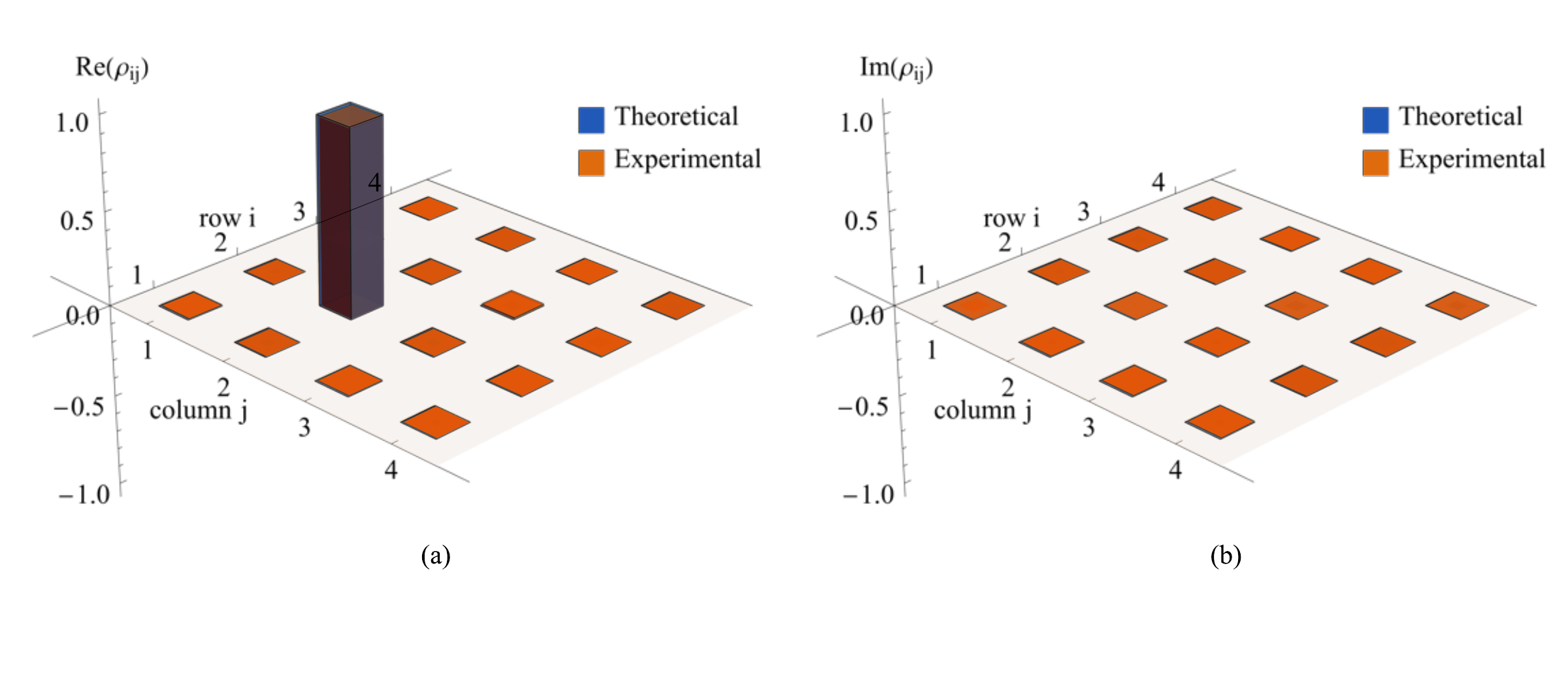}
  \caption{Real (a) and imaginary (b) parts of the density matrix for the state $\ket{\Psi_{4,3}}$. The blue (outer) bars correspond to theory, and the orange (nested) bars correspond to experiment.}
  \label{fig:A4_3}
\end{figure}

\subsection*{State $\ket{\Psi_{4,4}}$}

\noindent\textbf{Experimental density matrix:}
\begin{equation*}
{\rhoexp}_{44} = \begin{pmatrix}
  0.00625 & -0.0005 + 0.00125i & -0.00125 + 0.003i & 0.002 - 0.00225i \\
  -0.0005 - 0.00125i & 0.00425 & -0.001 - 0.00425i & -0.00075 - 0.001i \\
  -0.00125 - 0.003i & -0.001 + 0.00425i & -0.00475 & 0.0025 + 0.00775i \\
  0.002 + 0.00225i & -0.00075 + 0.001i & 0.0025 - 0.00775i & 0.9942
\end{pmatrix}
\end{equation*}
\noindent\textbf{Theoretical density matrix:}
\begin{equation*}
{\rhoideal}_{44} = \begin{pmatrix}
  0 & 0 & 0 & 0 \\
  0 & 0 & 0 & 0 \\
  0 & 0 & 0 & 0 \\
  0 & 0 & 0 & 1.
\end{pmatrix}
\end{equation*}
\noindent Purity: $0.99 \pm 0.03$; fidelity: $0.994 \pm 0.016$.

\begin{figure}[h!]
  \centering
  \includegraphics[width=\textwidth]{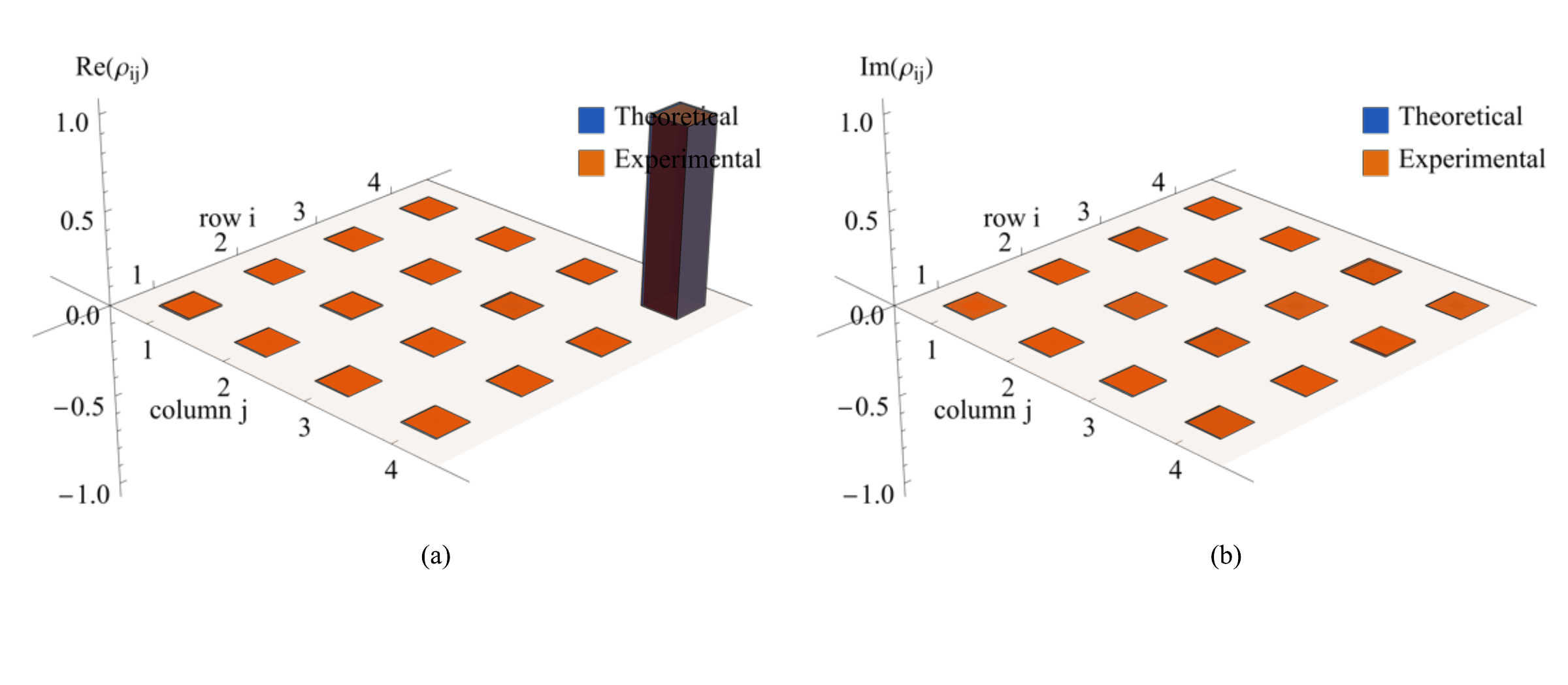}
  \caption{Real (a) and imaginary (b) parts of the density matrix for the state $\ket{\Psi_{4,4}}$. The blue (outer) bars correspond to theory, and the orange (nested) bars correspond to experiment.}
  \label{fig:A4_4}
\end{figure}

\subsection*{State $\ket{\Psi_{5,1}}$}

\noindent\textbf{Experimental density matrix:}
\begin{equation*}
{\rhoexp}_{51} = \begin{pmatrix}
  0.2465 & -0.009394 + 0.2463i & -0.2446 + 0.002949i & 0.007551 + 0.2436i \\
  -0.009394 - 0.2463i & 0.2539 & 0.007703 + 0.2528i & 0.2542 - 0.001518i \\
  -0.2446 - 0.002949i & 0.007703 - 0.2528i & 0.2564 & -0.007688 - 0.2485i \\
  0.007551 - 0.2436i & 0.2542 + 0.001518i & -0.007688 + 0.2485i & 0.2432
\end{pmatrix}
\end{equation*}
\noindent\textbf{Theoretical density matrix:}
\begin{equation*}
{\rhoideal}_{51} = \begin{pmatrix}
  0.25 & 0.25i & -0.25 & 0.25i \\
  -0.25i & 0.25 & 0.25i & 0.25 \\
  -0.25 & -0.25i & 0.25 & -0.25i \\
  -0.25i & 0.25 & 0.25i & 0.25
\end{pmatrix}
\end{equation*}
\noindent Purity: $0.99 \pm 0.02$; fidelity: $0.995 \pm 0.012$.

\begin{figure}[h!]
  \centering
  \includegraphics[width=\textwidth]{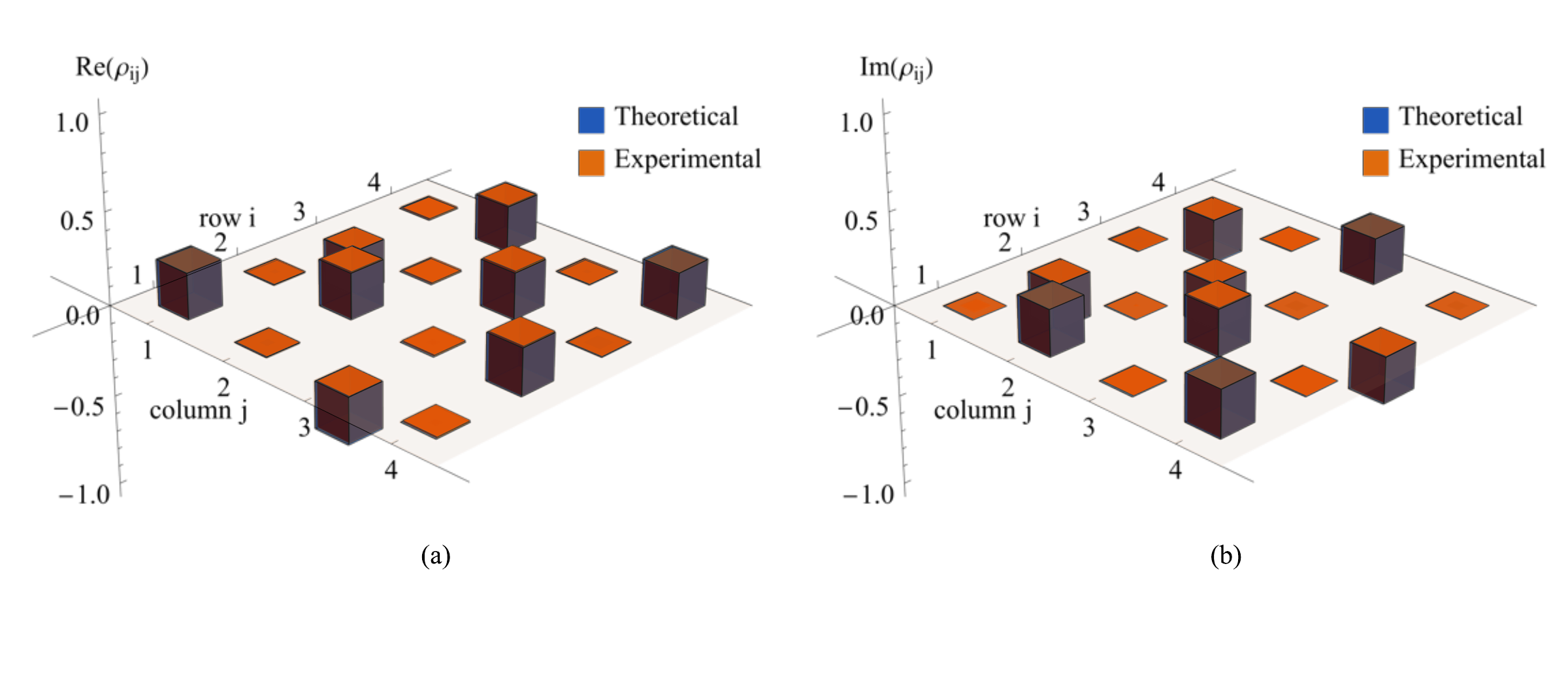}
  \caption{Real (a) and imaginary (b) parts of the density matrix for the state $\ket{\Psi_{5,1}}$. The blue (outer) bars correspond to theory, and the orange (nested) bars correspond to experiment.}
  \label{fig:A5_1}
\end{figure}

\subsection*{State $\ket{\Psi_{5,2}}$}

\noindent\textbf{Experimental density matrix:}
\begin{equation*}
{\rhoexp}_{52} = \begin{pmatrix}
  0.2393 & -0.006123 - 0.2446i & -0.2462 + 0.00436i & -0.003366 - 0.2465i \\
  -0.006123 + 0.2446i & 0.2473 & -0.003191 - 0.2527i & 0.2438 - 0.0008373i \\
  -0.2462 - 0.00436i & -0.003191 + 0.2527i & 0.2726 & 0.0006408 + 0.2479i \\
  -0.003366 + 0.2465i & 0.2438 + 0.0008373i & 0.0006408 - 0.2479i & 0.2408
\end{pmatrix}
\end{equation*}
\noindent\textbf{Theoretical density matrix:}
\begin{equation*}
{\rhoideal}_{52} = \begin{pmatrix}
  0.25 & -0.25i & -0.25 & -0.25i \\
  0.25i & 0.25 & -0.25i & 0.25 \\
  -0.25 & 0.25i & 0.25 & 0.25i \\
  0.25i & 0.25 & -0.25i & 0.25
\end{pmatrix}
\end{equation*}
\noindent Purity: $0.98 \pm 0.06$; fidelity: $0.99 \pm 0.03$.

\begin{figure}[h!]
  \centering
  \includegraphics[width=\textwidth]{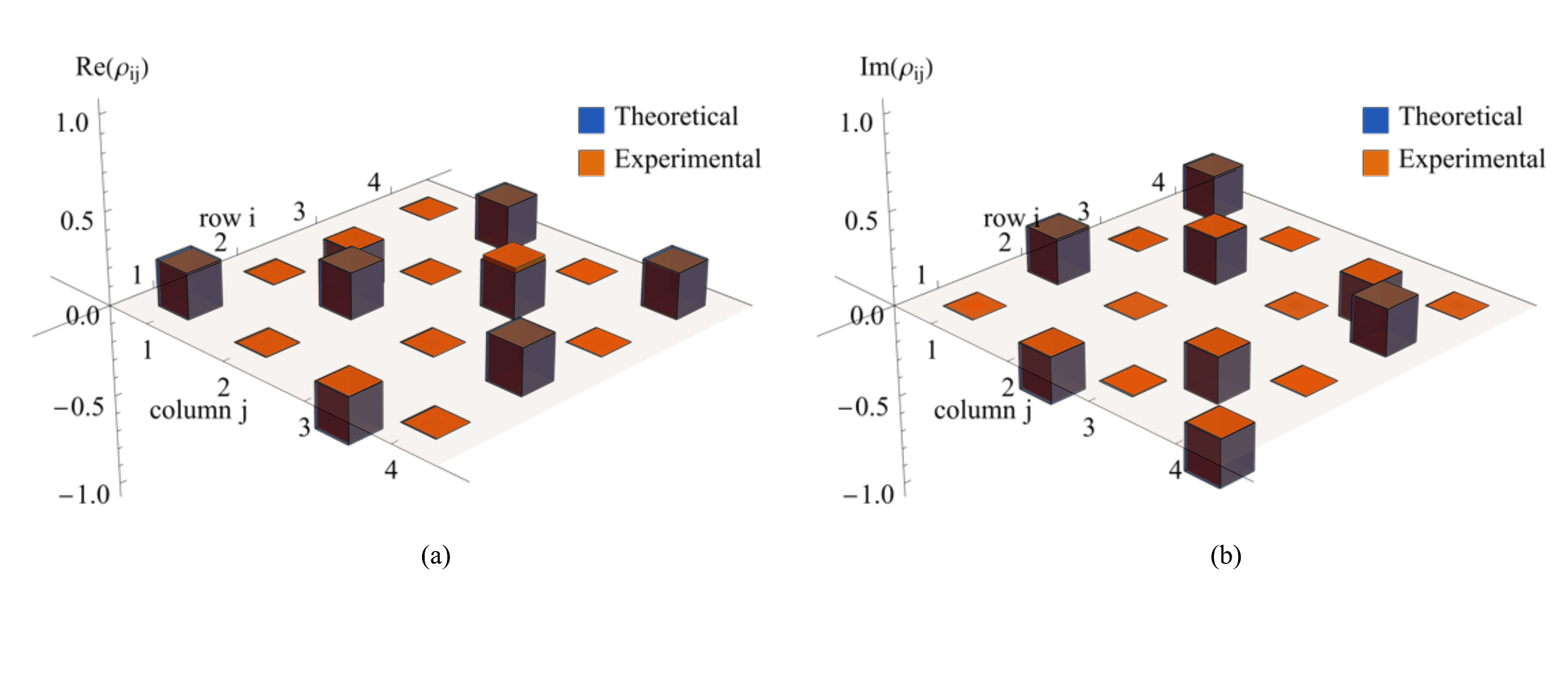}
  \caption{Real (a) and imaginary (b) parts of the density matrix for the state $\ket{\Psi_{5,2}}$. The blue (outer) bars correspond to theory, and the orange (nested) bars correspond to experiment.}
  \label{fig:A5_2}
\end{figure}

\subsection*{State $\ket{\Psi_{5,3}}$}

\noindent\textbf{Experimental density matrix:}
\begin{equation*}
{\rhoexp}_{53} = \begin{pmatrix}
  0.2407 & 0.009681 + 0.2673i & 0.2406 + 0.00852i & -0.005029 - 0.2512i \\
  0.009681 - 0.2673i & 0.2596 & -0.002947 - 0.2381i & -0.254 + 0.005768i \\
  0.2406 - 0.00852i & -0.002947 + 0.2381i & 0.2581 & 0.01948 - 0.2349i \\
  -0.005029 + 0.2512i & -0.254 - 0.005768i & 0.01948 + 0.2349i & 0.2416
\end{pmatrix}
\end{equation*}
\noindent\textbf{Theoretical density matrix:}
\begin{equation*}
{\rhoideal}_{53} = \begin{pmatrix}
  0.25 & 0.25i & 0.25 & -0.25i \\
  -0.25i & 0.25 & -0.25i & -0.25 \\
  0.25 & 0.25i & 0.25 & -0.25i \\
  0.25i & -0.25 & 0.25i & 0.25
\end{pmatrix}
\end{equation*}
\noindent Purity: $0.99 \pm 0.06$; fidelity: $0.99 \pm 0.03$.

\begin{figure}[h!]
  \centering
  \includegraphics[width=\textwidth]{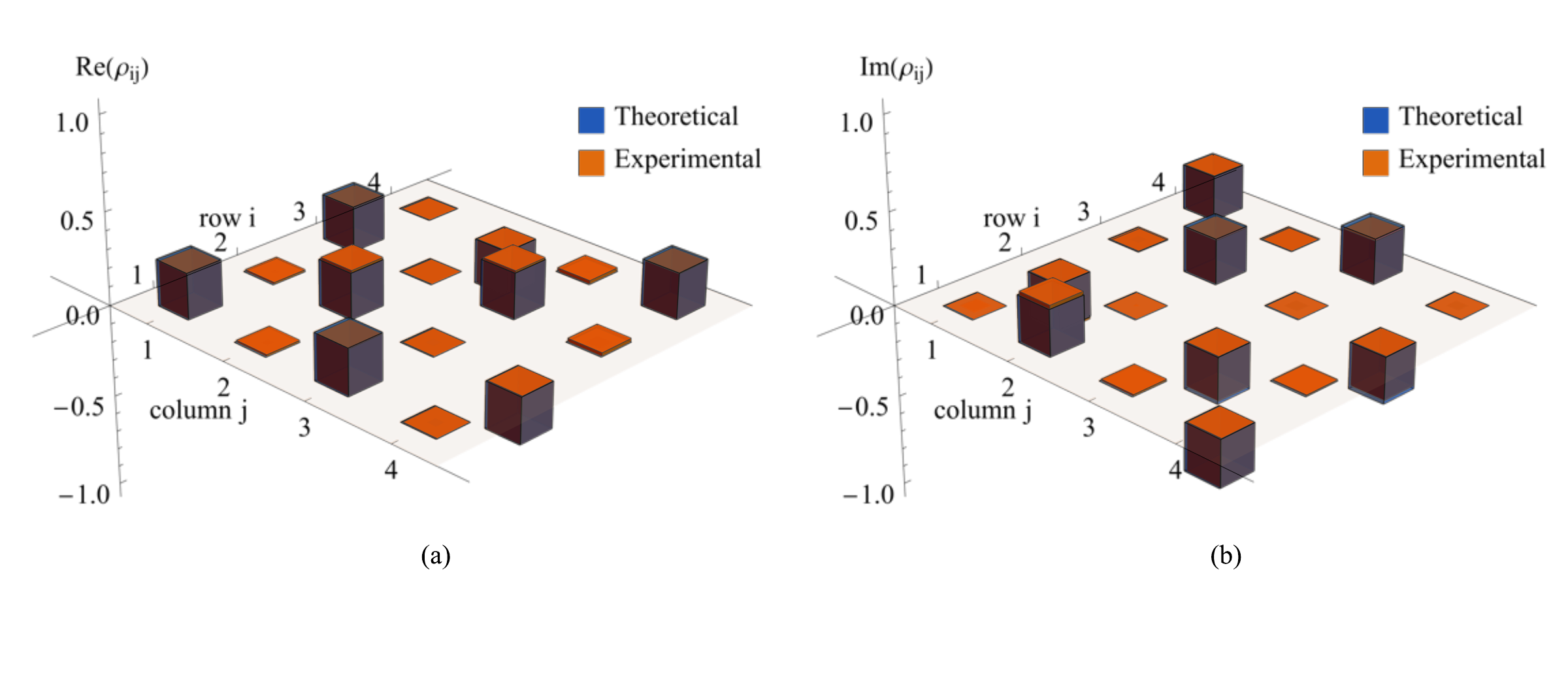}
  \caption{Real (a) and imaginary (b) parts of the density matrix for the state $\ket{\Psi_{5,3}}$. The blue (outer) bars correspond to theory, and the orange (nested) bars correspond to experiment.}
  \label{fig:A5_3}
\end{figure}

\subsection*{State $\ket{\Psi_{5,4}}$}

\noindent\textbf{Experimental density matrix:}
\begin{equation*}
{\rhoexp}_{54} = \begin{pmatrix}
  0.2691 & -0.005276 - 0.2438i & 0.2539 - 0.0003079i & -0.006815 + 0.2342i \\
  -0.005276 + 0.2438i & 0.2296 & 0.003073 + 0.2505i & -0.2556 + 0.007275i \\
  0.2539 + 0.0003079i & 0.003073 - 0.2505i & 0.2531 & -0.01898 + 0.2375i \\
  -0.006815 - 0.2342i & -0.2556 - 0.007275i & -0.01898 - 0.2375i & 0.2482
\end{pmatrix}
\end{equation*}
\noindent\textbf{Theoretical density matrix:}
\begin{equation*}
{\rhoideal}_{54} = \begin{pmatrix}
  0.25 & -0.25i & 0.25 & 0.25i \\
  0.25i & 0.25 & 0.25i & -0.25 \\
  0.25 & -0.25i & 0.25 & 0.25i \\
  -0.25i & -0.25 & -0.25i & 0.25
\end{pmatrix}
\end{equation*}
\noindent Purity: $0.98 \pm 0.04$; fidelity: $0.99 \pm 0.02$.

\begin{figure}[h!]
  \centering
  \includegraphics[width=\textwidth]{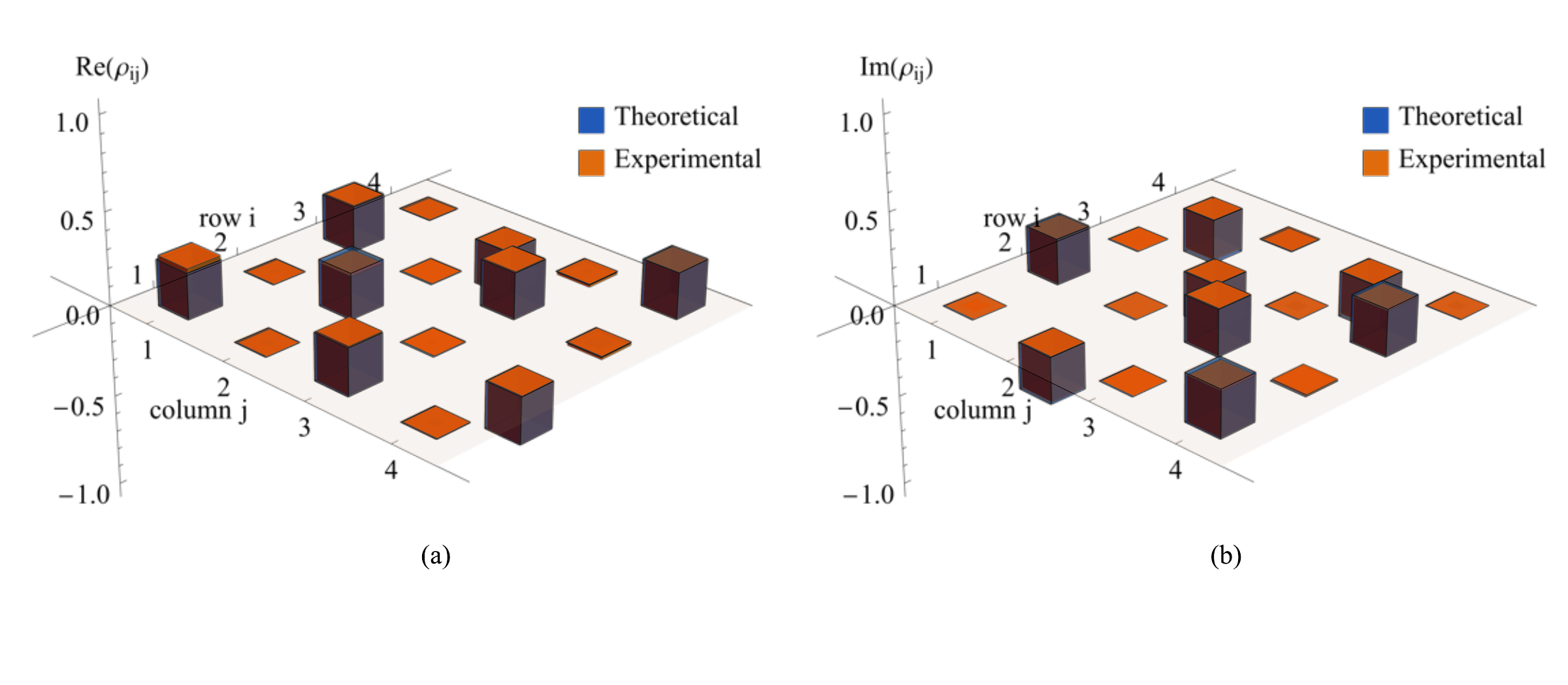}
  \caption{Real (a) and imaginary (b) parts of the density matrix for the state $\ket{\Psi_{5,4}}$. The blue (outer) bars correspond to theory, and the orange (nested) bars correspond to experiment.}
  \label{fig:A5_4}
\end{figure}

\end{document}